\input harvmac
\input amssym.tex

%title
\def\title#1#2#3{\nopagenumbers\abstractfont\hsize=\hstitle
\rightline{hep-th/9708004}
\rightline{ITFA-#1}
\bigskip
\vskip 0.7in\centerline{\titlefont #2}\abstractfont\vskip .2in
\centerline{{\titlefont#3}}
\vskip 0.3in\pageno=0}
%name and address
\def\authors{ \centerline{
{\bf F.A. ~Bais}\footnote{$^{\dagger}$}
{e-mail: {\tt bais@phys.uva.nl}} and 
{\bf B.J. ~Schroers}\footnote{$^{\ddagger}$}
{e-mail: {\tt schroers@phys.uva.nl} } }
\bigskip
\centerline{
Instituut  voor  Theoretische  Fysica}
\centerline{Universiteit van  Amsterdam} 
\centerline{
Valckenierstraat 65, 1018 XE Amsterdam}
\centerline{The Netherlands}
\bigskip\bigskip}
%abstract
\def\abs#1{\centerline{{\bf Abstract}}\vskip 0.2in
\baselineskip 12pt
{#1}}
\def\ack{\bigbreak\bigskip\smallskip\centerline{{\bf Acknowledgements}}
\nobreak}

%The bold versions of the lower-case Greek letters.

%\font\tenbifull=cmmib10 % bold math italic
%\font\tenbimed=cmmib10 scaled 800
%\font\tenbismall=cmmib10 scaled 666
%\textfont10=\tenbifull \scriptfont10=\tenbimed
%\scriptscriptfont10=\tenbismall
%\def\bmit{\fam10 }
%

%\def\bal{{\fam=9{\mathchar"710B } }}
%\def\bett{{\fam=10{\mathchar"710C } }}
%\def\bphi{{\fam=9{\mathchar"711F} }}
%\def\bpi{{\fam=9{\mathchar"7119 } }}
\def\bett{\vec{\beta} }

\def\cd{\!\cdot\!}

\def\bCP{{\hbox{\bf CP}}}
\def\bC{{\hbox{\bf C}}}
\def\bk{{\hbox{\bf k}}}
\def\ba{{\hbox{\bf a}}}
\def\br{{\hbox{\bf r}}}
\def\bI{{\hbox{\bf I}}}
\def\bH{\vec H}

\def\bg{\vec g}
\def\bh{\vec h}
\def\bZ{{\hbox{\bf Z}}}
\def\bR{{\hbox{\bf R}}}

 \def\mapd{\Big\downarrow
  \rlap{$\pi^k $} }
\def\pmatrix#1{\left(\matrix{#1}\right)}
\def\semidir{\ltimes }
%\def\semidir{(\times}

% Euler Fonts
%\font\teneufm=eufm10
%\font\seveneufm=eufm7
%\font\fiveeufm=eufm5
%\newfam\eufmfam
%\textfont\eufmfam=\teneufm
%\scriptfont\eufmfam=\seveneufm
%\scriptscriptfont\eufmfam=\fiveeufm
%\def\eufm#1{{\fam\eufmfam\relax#1}}

%\font\teneusm=eusm10
%\font\seveneusm=eusm7
%\font\fiveeusm=eusm5
%\newfam\eusmfam
%\textfont\eusmfam=\teneusm
%\scriptfont\eusmfam=\seveneusm
%\scriptscriptfont\eusmfam=\fiveeusm
%\def\eusm#1{{\fam\eusmfam\relax#1}}
 
%\font\tenmsx=msam10
%\font\sevenmsx=msam7
%\font\fivemsx=msam5
%\font\tenmsy=msbm10
%\font\sevenmsy=msbm7
%\font\fivemsy=msbm5
%\newfam\msafam
%\newfam\msbfam
%\textfont\msafam=\tenmsx  \scriptfont\msafam=\sevenmsx
%  \scriptscriptfont\msafam=\fivemsx
%\textfont\msbfam=\tenmsy  \scriptfont\msbfam=\sevenmsy
%  \scriptscriptfont\msbfam=\fivemsy
%\def\msam#1{{\fam\msafam\relax#1}}
%\def\msbm#1{{\fam\msbfam\relax#1}}

%----------------
\lref\HP{G. 't Hooft, ``Magnetic monopoles in unified 
gauge theories'', Nucl. Phys. {\bf B 79} (1974) 276
\semi
A.M. Polyakov, ``Particle spectrum in quantum field theory''
JETP Letters {\bf 20} (1974) 194.
}
\lref\MO{C. Montonen and D.Olive, ``Magnetic monopoles 
as gauge particles ?'', Phys. Lett. {\bf  B 72 }
(1977) 117.
}
\lref\GNO{ P. Goddard, J. Nuyts and D. Olive, ``Gauge theories and
magnetic charge'', Nucl. Phys. {\bf B 125} (1977) 1.
}
\lref\VaWi{C. Vafa and E. Witten, ``A strong coupling test of 
S-duality'', Nucl. Phys. {\bf B 431} (1994) 3. 
}
\lref\EW{F. Englert and P. Windey, ``Quantization condition for 
't Hooft monopoles in compact simple Lie groups'', Phys. Rev. {\bf D 14}
(1976) 2728.
}
\lref\Corrigan{E. Corrigan, ``Static non-abelian forces and the 
permutation group'', Phys. Lett. {\bf B 82 } (1979) 407.
}
\lref\SikWe{P. Sikivie and N. Weiss, ``Classical Yang-Mills theory
in the presence of external sources'', Phys. Rev. {\bf D 18} (1978) 3809
\semi
``Static sources in classical Yang-Mills theory'', Phys. Rev. {\bf D 20}
487.
}
\lref\Godol{P. Goddard and D. Olive, 
``Charge Quantization in theories with an adjoint representation
Higgs mechanism'', Nucl. Phys. {\bf B191} 511 \semi
``The magnetic charges of stable self-dual monopoles'' Nucl. Phys.
 {\bf B 191}(1981) 528.
}
\lref\Andd{A.S. Dancer, ``Nahm's equation and Hyperk\"ahler Geometry'',
Commun. Math. Phys.  {\bf  158} (1993) 545.
}
\lref\Anddd{ A.S. Dancer, ``A family of Hyperk\"ahler Manifolds'', 
Quart. Jour. Math.  {\bf 45} (1994)  463.  
}
\lref\And{ A.S. Dancer, ``Nahm data and $SU(3)$ monopoles'',
 Nonlinearity {\bf  5} (1992) 1355.
}
\lref\DL{A.S. Dancer and R.A. Leese, ``Dynamics of $SU(3)$ 
monopoles'', Proc. Roy. Soc. Lond. 
{\bf A440} (1993) 421.
}
\lref\DLL{A.S. Dancer and R.A. Leese, ``A numerical study of $SU(3)$ 
charge-two monopoles with minimal symmetry breaking'', Phys. Lett.
{\bf  B 390} (1997) 252.
}
\lref\Rich{R.S. Ward, 
``Magnetic monopoles with gauge group $SU(3)$ broken to 
$U(2)$'', Phys. Lett. {\bf B 107} (1981) 281.
}
\lref\Richh{ R.S. Ward, 
``Deformations of the embedding of the $SU(2)$ monopole
solution in $SU(3)$'', Commun. Math. Phys. {\bf 86} (1982) 437.
}
\lref\FHP{P. Forgacs, Z. Horvath and L. Palla,
``On the construction of axially 
symmetric $SU(n)$  monopoles'',
Nucl. Phys. {\bf B 221} (1983) 235.
}
\lref\ansat{A. Chakrabarti, ``On classical solutions of $SU(3)$ gauge 
field equations'', Ann. Inst. H.  Poincar\'e {\bf 23} (1975)
235 
\semi
W.J. Marciano and Heinz Pagels,  ``Classical $SU(3)$ gauge theory and magnetic
monopoles'', Phys. Rev. {\bf D 12}, (1975) 1093
\semi
Z. Horv\'ath and L. Palla, ``Dyons in classical $SU(3)$ gauge theory
and a new topologically conserved quantity'', Phys. Rev. {\bf D 14}
(1976) 1711
\semi
E. Corrigan, D.I. Olive, D.B. Fairlie and J. Nuyts, 
``Magnetic monopoles in $SU(3)$ gauge theory'', Nucl. Phys. {\bf B 106}
(1976) 475
}
\lref\COO{E. Corrigan and D. Olive, ``Colour and magnetic monopoles'',
Nucl. Phys. {\bf  B 110} (1976) 237.
}
\lref\Sander{F.A. Bais, ``Charge-monopole duality in spontaneously
broken gauge theories'', Phys. Rev. {\bf D 18} (1978) 1206.
}
\lref\BW{F.A. Bais and H.A. Weldon, ``Exact Monopole Solutions 
in $SU(n)$ gauge theory'' Phys. Rev. Lett. {\bf 41}
(1978) 601 \semi
D. Wilkinson and   F.A. Bais, ``Exact $SU(N)$ monopole
solutions with spherical symmetry' , Phys. Rev. {\bf D 19}
(1979) 2410.
}
\lref\Wein{E.J. Weinberg, ``Fundamental monopoles and multimonopole
solutions for arbitrary simple gauge groups'', Nucl. Phys. {\bf B 167}
(1980) 500.
}
\lref\Taubes{C.H. Taubes, ``Surface integrals and monopole charges
in non-abelian gauge theories'', Commun. Math. Phys. {\bf 81} (1981)
299.
}
\lref\Taubess{C.H. Taubes, ``The existence of multi-monopole solutions
to the non-abelian Yang-Mills-Higgs equations and arbitrary gauge
groups'', Commun. Math. Phys. {\bf 80}, (1981)  343.
}
\lref\Abou{A. Abouelsaood, ``Chromodyons and equivariant gauge 
transformations'', Phys. Lett. {\bf B  125} (1983) 467.
}
\lref\Abouu{A. Abouelsaood, ``Are there chromodyons?'', Nucl. Phys. 
{\bf B 226} (1983) 309.
}
\lref\NC{P. Nelson and S. Coleman, ``What becomes of global color ?'',
Nucl. Phys. {\bf B 237} (1984) 1.
}
\lref\NM{P.N. Nelson and A. Manohar, ``Global colour is not
always defined'', Phys. Rev. Lett. {\bf 50} (1983) 943.
} 
\lref\BMMNSZ{
A. Balachandran {\it et al.}, ``Monopole topology and the problem of color'',
 Phys. Rev. Lett {\bf 50} (1983) 1553.
}
\lref\Swansea{N. Dorey, C. Fraser, T.J. Hollowood and M.A.C. Kneip,
``Non-abelian duality in N=4 supersymmetric gauge theories'', 
{\tt hep-th/9512116}
 \semi
``S-duality in N=4 supersymmetric gauge theories with arbitrary
gauge group'',Phys. Lett.{\bf  B 383} (1996) 422.
}
\lref\Murray{M. Murray, ``Stratifying monopoles and rational maps'',
Commun. Math. Phys. {\bf 125} (1989) 661.
}
\lref\Jarvis{S. Jarvis, ``Euclidean monopoles and rational maps
'', Oxford University preprint 1997, submitted to Proc. London
Math. Soc. . 
}
\lref\Bowman{M.C. Bowman, ``Parameter counting for self-dual monopoles'',
Phys. Rev. {\bf D 32} (1985) 1569.
}
\lref\athi{M.F. Atiyah and N. Hitchin, {\it The geometry and dynamics
of magnetic monopoles}, Princeton University Press, New Jersey 1988.
}
\lref\atmu{M.F. Atiyah, ``Instantons in two and four dimensions'',
Commun. Math. Phys.  {\bf 93} (1984) 437
\semi
M.F. Atiyah, ``Magnetic monopoles in hyperbolic space'', In ``Vector
bundles on Algebraic Varieties'', Oxford University Press, 1987.
}
\lref\donaldson{ S.K. Donaldson, ``Nahm's equation and the classification
of monopoles'', Commun. Math. Phys. {\bf 96} (1984) 387.
}
\lref\sen{A. Sen, ``Dyon-monopole bound states, self-dual harmonic
forms on the multi-monopole moduli space, and $SL(2,\bZ)$ invariance 
in string theory'', Phys. Lett. {\bf B 329 } (1994) 217. 
}
\lref\segal{G. Segal and A. Selby, ``Cohomology of spaces of monopoles'',
Commun. Math. Phys. {\bf 177} (1996) 775.
}
\lref\abduality{ J. Gauntlett and D. Lowe,
``Dyons and S-duality in N=4 supersymmetric gauge theory'',
 Nucl. Phys.{\bf  B472} (1996) 194 
\semi
 K. Lee, E.J.  Weinberg and  P. Yi, ``Electromagnetic duality and SU(3)  
monopoles'' Phys. Lett.{\bf B  376 } (1996) 97
\semi
G.W. Gibbons, ``The Sen conjecture for fundamental monopoles of distinct
type'', Phys. Lett. {\bf B 382 } (1996) 53. 
}
\lref\wardcp{R.S. Ward, ``Lumps in the $\bCP^1$-model in (2+1)
dimensions'', Phys. Lett. {\bf B 158 } (1985) 424.
}
\lref\LWY{K. Lee, E.J. Weinberg and P. Yi, ``Massive and massless monopoles
with nonabelian magnetic charges'',Phys. Rev. {\bf D 54}
(1996) 6351.
}
\lref\irwin{ P. Irwin, ``SU(3) monopoles and their fields'', DAMTP preprint
97-33$,\,\,\,$  {\tt hep-th/9704153}.
}
\lref\MS{ N.S. Manton and T.M. Samols ``Radiation from monopole scattering'',
Phys.Lett. {\bf B 215 } (1988) 559. 
}
\lref\GM{G.W. Gibbons and N.S. Manton, 
 ``Classical and Quantum Dynamics of BPS monopoles'', Nucl. Phys.
{\bf B 274} (1986) 183.
}
\lref\SaBe{F.A. Bais and B.J. Schroers, 
``Non-abelian dyons and S-duality'',
in preparation.
}
\lref\indrep{A.O. Barut and R. Raczka, {\it Theory of group representations
and applications}, PWN, Warszawa 1977.
}
\lref\cgcoeff{J.S. Rno, ``Clebsch-Gordan coefficients and special functions
related to the Euclidean group in three-space'', J. Math. Phys. {\bf 15}
(1974) 2042
\semi
N.P. Landsman, ``Non-shell unstable particles in thermal field theory'',
Ann. Phys. {\bf 186} (1988) 141.
}
\lref\berni{B.J. Schroers, ``Quantum scattering of BPS monopoles 
at low energy'', Nucl. Phys. {\bf B 367} (1991) 177.
}
\lref\susymo{J. P. Gauntlett, ``Low-energy dynamics of N=2 supersymmetric
monopoles'', Nucl. Phys. {\bf B 411} (1994) 443
\semi
J. D. Blum, ``Supersymmetric quantum mechanics of monopoles in
N=4 Yang-Mills Theory'', Phys. Lett. {\bf B  333 } (1994) 92.
}
\lref\threerus{D.A. Varshalovich, A.N. Moskalev and V.K. Khersonskii, 
{\it Quantum Theory of Angular Momentum}, World Scientific, Singapore 1988.
}
\lref\BDP{F.A. Bais, P. van Driel and M. de Wild Propitius, ``Quantum
 symmetries in discrete gauge theories'', Phys. Lett. {\bf B 280} (1992) 63
\semi
 M. de Wild Propitius and  F.A. Bais, ``Discrete gauge theories'',
Proc. CRM-CAP Summer School ``Particles and fields '94'', Banff, Springer
Verlag (1996),  {\tt  hep-th/9511201}.
}
\lref\MK{T.H. Koornwinder and N.M. Muller, ``The quantum double of a 
(locally) compact group'', J. of Lie Theory {\bf 7} (1997) 101;
``The quantum double of $SU(2)$'',  in preparation\semi 
F.A. Bais, T.H. Koornwinder and N.M. Muller 
``Decomposition of tensor products of the quantum double of a 
compact group'', in preparation.
}
\lref\IW{E. In\"on\"u and E.P. Wigner, ``On the contraction of groups
and their representations'', Proc. Nat. Acad. Sci. (US) {\bf 39} (1953)
510.
}
\lref\alvar{M. Alvarez, ``Physical states of dyons'',  SWAT-97-156,
 {\tt hep-th/9707117}.
}
\lref\braner{R.A. Brandt and F. Neri, ``Stability analysis for singular
non-abelian magnetic monopoles'', Nucl. Phys. {\bf B 161} (1979) 253.
}
\lref\lozano{Y. Lozano, ``Duality and canonical transformations'',
Talk at Argonne Duality Institute, June 1996, {\tt hep-th/9610024}.
}
%---------------------------------

\title{97-28}
{Quantisation of  monopoles with}{ non-abelian magnetic charge}

\authors
 
\abs
\noindent 
Magnetic monopoles in Yang-Mills-Higgs theory with a non-abelian 
unbroken gauge group  are classified by holomorphic charges 
in addition to the topological charges familiar from the abelian case. 
As a result the moduli spaces of monopoles of given topological charge 
are stratified according to the holomorphic charges. Here the physical 
consequences of the stratification are explored in the case where 
the gauge group $SU(3)$ is broken to $U(2)$. The description due to 
A. Dancer of the moduli space of charge two monopoles is reviewed and 
interpreted physically in terms of non-abelian magnetic dipole moments. 
Semi-classical quantisation leads to dyonic states which are labelled  
by a magnetic charge and a representation of the subgroup  of $U(2)$ 
which leaves the  magnetic charge invariant (centraliser subgroup). 
A key result of this paper is that these states fall into representations 
of the  semi-direct product $U(2) \semidir \bR^4$. The combination rules 
(Clebsch-Gordan coefficients) of dyonic states can thus be  deduced.
Electric-magnetic duality properties of the theory are discussed in 
the light of our results, and supersymmetric dyonic BPS states 
which fill the $SL(2,\bZ)$-orbit of the basic massive $W$-bosons are 
found.

\hbox{}

\centerline{\it to appear in Nuclear Physics B}

%\Date{July  1997}

%\draftmode
\baselineskip 16pt
\newskip\normalparskip
\normalparskip = 4pt plus 1.2pt minus 0.6pt
\parskip = \normalparskip

\newsec{Introduction}
The study of  magnetic monopole solutions in  spontaneously broken 
gauge theories,  sparked off more than twenty years ago by 't Hooft's 
and Polyakov's  discovery of the eponymous monopole solution
in SU(2) Yang-Mills-Higgs theory \HP, has progressed from  
the classification
and in some cases explicit construction of monopoles via the 
description of the spaces of solutions (the moduli spaces)  to, 
more recently, the discussion of 
classical and quantised  dynamics of monopoles. Perhaps not
surprisingly, most progress has been made in the theory originally
considered by 't Hooft and Polyakov where, in a special limit called the 
BPS limit,  the understanding of the geometry of the classical
moduli spaces could  be used rigorously to  establish
the existence of infinitely many (supersymmetric) quantum bound
states of magnetic monopoles \sen \segal. These bound states,
which are typically dyonic,
are related to the electrically charged   $W$-bosons of the
theory via electric-magnetic duality or S-duality and their
existence confirms the possibility of formulating Yang-Mills
theory in infinitely many equivalent ways, all related by 
S-duality and each having different particles as fundamental
degrees of freedom.

As far as the spectrum of magnetically  and electrically
charged particles  is  concerned the step from 
  Yang-Mills-Higgs
theory with gauge group $SU(2)$  broken to $U(1)$ 
to a general gauge group $G$  broken to some subgroup $H$
is not   of great qualitative difficulty provided the group
$H$  is abelian. Indeed, in the case $G=SU(N)$ and $H=U(1)^{N-1}$
(a maximal torus of $G$) much is known about monopole solutions, 
their moduli spaces, their quantum bound states and the action of 
electric-magnetic duality,  even quantitatively \abduality. 
When $H$ is non-abelian,
however, it was already pointed out more than ten years ago
  by a number of authors  that  qualitatively 
new problems arise when attempting
physically  to interpret parameters of multimonopole
solutions and when discussing 
dyonic excitations of monopoles. The many interesting physical 
problems unearthed  by these discussions, however, were largely
ignored in the more mathematical treatment  of monopoles
and their moduli spaces in recent years. In attempting the 
final of the steps outlined above, that of trying to apply
the mathematical understanding of monopole moduli spaces to  
the study of  classical and quantum dynamics of monopoles, 
these problems  naturally return. In particular they beset
any attempt of  understanding  electric-magnetic duality
in these theories.  Let us therefore briefly review
the issues involved.

The allowed values of magnetic charges in 
 non-abelian gauge theories 
are restricted by the generalised  Dirac quantisation condition see
\EW, \GNO, and also \COO. A naive
application of this condition leads to the representation
of the allowed magnetic 
charges as points on the dual of the weight lattice
of the full gauge group $G$. Thus, if we denote the rank of 
$G$ by $R$, magnetic charges correspond
to vectors in a $R$-dimensional lattice,
usually called magnetic weight vectors. This general condition
does not reflect  the specifics of the symmetry  breaking. 
On the other hand there is a topological classification of 
magnetic charges which depends crucially on  the breaking of 
$G$ down to the exact residual gauge group $H$. The topological
charges are elements of  the second homotopy group $\Pi_2(G/H)$
($=\Pi_1(H)$  if $G$ is simply connected) and thus depend strongly
on the connectedness of $H$. If one breaks the symmetry  
to a maximal torus $H \cong U(1)^R$  of $G$ (maximal symmetry breaking)
 the classification
resulting from the Dirac condition agrees with the topological
classification: all $R$ components of the magnetic weight vectors
are topologically conserved. Now consider the degenerate  situation
where one does not break maximally and the exact group $H$
is non-abelian. The magnetic weight vectors now have more components
than the number of topologically conserved charges, raising the 
question of which - if any - relevance  the remaining components
(called the non-abelian components)
might have. The answer is rather subtle and depends on additional
assumptions. If one carries the Brandt-Neri reasoning \braner\  over
to the unbroken group $H$ one might expect that  configurations
whose magnetic weight vectors have non-abelian components
will decay to configurations
whose non-abelian components are, in a suitable sense, minimal.
There is no topological  obstruction to doing so.
However, if one considers the theory in the BPS limit, shedding
non-abelian charge in this way does not lower the energy.
Mathematical analysis has revealed that there are neutrally
stable solutions which have magnetic weights with non-minimal
non-abelian components (restricted to lie in a certain range).
In this situation the magnetic weights regain some of their 
glory. Mathematically they characterise holomorphic properties
of the solution and they (or more precisely certain equivalence
classes of them) are called holomorphic charges
in the mathematical literature. Thus one may say that in 
general the magnetic weight vectors have topological and 
holomorphic components.

To illustrate the general discussion consider the simplest
non-trivial example where $G=SU(3)$ and $H=U(1)\times U(1)$
or $H=U(2)$ for, respectively,
 maximal or  minimal symmetry breaking.
In Fig.~1 we show the magnetic weight lattice, spanned by the two
simple roots $\bett_1$ and $\bett_2$, together with the direction $\bh$
of the Higgs field which determines the symmetry breaking pattern.
For each point in the weight lattice we indicate
the dimensionality of the corresponding space of monopole
solutions, called moduli space and to be defined 
more precisely in the  main part of the paper. In the case
of maximal symmetry breaking there are non-trivial moduli spaces
for all positive magnetic weights (there are also moduli
spaces for negative magnetic weights, containing anti-monopoles,
but we do not consider these here). In the 
case of minimal symmetry breaking non-trivial moduli spaces only
occur inside the cone shown in Fig.~1.b, where the topological
component of the magnetic weight vectors is plotted vertically and 
the holomorphic component horizontally. The geometrical structure
of these spaces is  intricate. Each topological sector corresponds
to a connected moduli space which consists of different strata
of varying dimensions, with each stratum labelled by an (integer)
 holomorphic charge. There are one-parameter families of solutions
where the holomorphic charge jumps from one discrete value to the next
as the parameter varies continuously. These mathematical facts
have so far not been interpreted physically. What is the physical
meaning of the parameters which appear and disappear along the 
journey through the moduli space?

\vskip .3in
\input epsf 
\epsfxsize=16truecm
\centerline{
\epsfbox{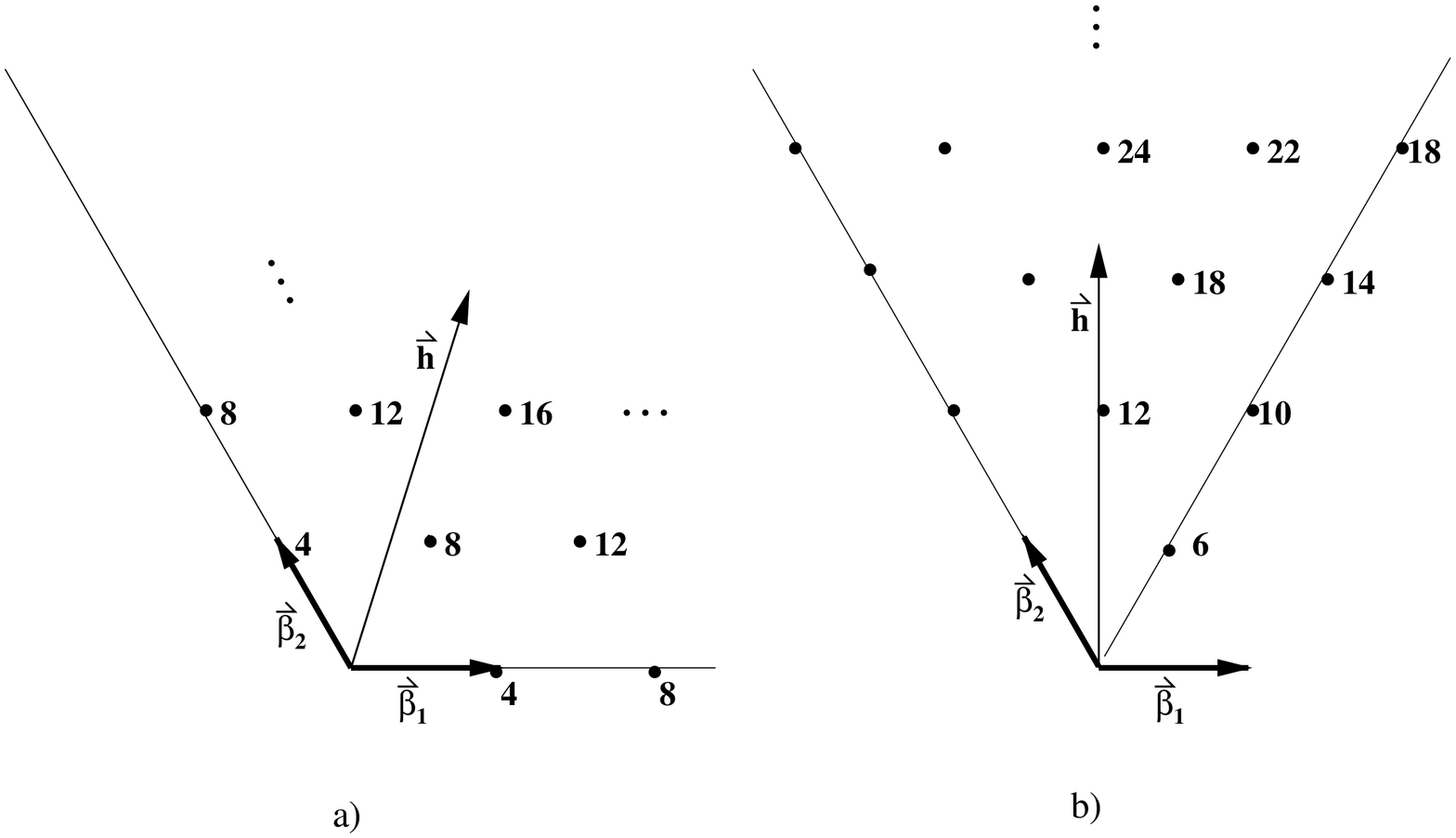}
}
\bigskip
{\footskip14pt plus 1pt minus 1pt \footnotefont
\centerline{\bf Fig.~1 }
\noindent a) { \it  Moduli spaces and their dimensions 
for $SU(3)$ monopoles with  maximal symmetry breaking }
\noindent b) {\it Strata of moduli spaces and their dimensions
for $SU(3)$ monopoles with minimal symmetry breaking}
}
\bigskip

There is a related  question which has attracted attention for
a long time but which has never been resolved satisfactorily.
This is  the question of  how the exact symmetry group $H$ is realised
in the various magnetic sectors.
If the magnetic weight vector has non-abelian components
it is not invariant under  the action of $H$ but 
instead  sweeps out an orbit, which
 we call the magnetic orbit in this paper.  All magnetic weight vectors on
such a magnetic  orbit (which includes the Weyl orbit
of the magnetic weight vector) are allowed
by the Dirac quantisation condition. In a given theory, however, different
sorts of orbits arise. In the minimally broken $SU(3)$  theory,
 the magnetic weight vectors on the vertical axis are invariant 
under the $U(2)$ action, but all other magnetic weight vectors
sweep out orbits which are two-spheres of quantised
radii  in the Lie-algebra of $SU(3)$. It is well known that 
``electric'' excitations of a monopole with given magnetic 
weight do not, as one might naively expect, fall into representations
of the exact group $H$ but  only form representations of the centraliser
subgroup  of the magnetic weight vector. This implies that 
the semi-classical dyon spectrum in a gauge theory with 
 a non-abelian unbroken gauge group $H$  displays  an intricate
interplay between  magnetic and electric quantum numbers.
In the $SU(3)$ example we see that  electric 
excitations of monopoles  with magnetic
weight vectors on the vertical axis  form $U(2)$ representations
while for the other magnetic weights the electric excitations
only carry representations of $U(1)\times U(1)$. Moreover, in
the latter case one of the $U(1)$  groups depends on the point
of the magnetic orbit to which the magnetic weight vector belongs.
In other words monopoles in this sector may be charged with respect
to different $U(1)$ subgroups. Clearly these matters have to be 
reconciled at the quantum level, where one expects a single
algebraic framework which allows for a unified characterisation
of both magnetic  and electric properties of the states, and
which furthermore  allows one to combine the various sectors
in some sort of tensor product calculus.
Finally, such a framework can be expected to be a crucial 
ingredient in a discussion of electric-magnetic duality properties
in gauge theories with non-abelian unbroken gauge group.

Having reviewed the physical problems we address in this paper,
we can now outline our strategy for tackling them by giving 
the plan of this  paper.
We confine  attention to  monopoles in $SU(3)$ Yang-Mills-Higgs 
theory.
The generalisation of arguments to   general gauge groups and 
symmetry breaking patterns will be presented in a separate 
paper \SaBe.
 
We begin  in Sect.~2 with a  review of classical monopole
solutions in $SU(3)$ Yang-Mills-Higgs theory. Then we go on,
in Sect.~3,
to define classical configuration spaces and the moduli spaces of 
monopole solutions of  the Bogomol'nyi equation.
 We review the stratification of the moduli spaces and discuss
some of their geometrical properties. Monopole solutions in 
the so-called smallest stratum can be obtained by embedding
$SU(2)$ monopole solutions in the $SU(3)$ theory, and this
is described in Sect.~4. We then take a break from the
mathematics of moduli spaces and in two short sections we 
formulate the two physical problems reviewed above - that of 
implementing the exact  $U(2)$ symmetry  and that of 
physically understanding the monopole moduli  - in  sharper
mathematical language.
For the rest of the paper we  focus on monopoles of topological
charge one and two. We review  the rational map description
of monopoles  
in Sect.~7 and in Sect.~8 we describe the work 
of Andrew  Dancer who investigated
the 12-dimensional large stratum of the charge  two monopole moduli
space  in great detail. 
The physical interpretation of Dancer's moduli space is tricky
because the 
mathematically most convenient  description 
of the space is  in terms of  objects such 
as rational maps or Nahm data which are  
 related to the actual monopole fields via mathematically
complicated transformations. Using a combination 
of the different descriptions  we are  able to  give a physical
interpretation of  the moduli of charge two monopoles in Sect.~9.
It had been noted long ago that the interpretation of the 
parameters  in terms of the zero-modes of two individual monopoles
is problematic and we will see that individual monopoles 
indeed have new zero-modes when combined into a multi-monopole
configuration. We argue that these new zero-modes are related 
to the possibility of the individual monopoles having non-vanishing
dipole moments when part of  a multimonopole configuration.

Sect.~10 is the key section of our paper. Here we  present a 
detailed study  of semi-classical dyonic excitations 
of monopoles.  Dyonic quantum states
are realised as wavefunctions on the strata of  the moduli space
and have to satisfy
certain superselection rules. The dependence of   
the $U(2)$ action on the moduli space  on the 
strata feeds trough to the  dyonic wavefunctions in just 
the expected way:
In the large stratum the $U(2)$ action  is free and differentiable
and as a result the monopoles 
carry full $U(2)$ representations. In all the other
strata, however, the non-abelian magnetic charge obstructs the $U(2)$
action  and the monopoles
only carry representations of the   $U(1)\times U(1)$ centraliser
subgroup 
of the non-abelian magnetic charge.
Remarkably we find that   general dyonic states 
may be interpreted as   elements of representations
of the semi-direct product of $U(2) \semidir \bR^4$.
Although this group does not act on the moduli spaces, it
does have a natural action on wavefunctions on the moduli
spaces. Thus quantum states, realised as wavefunctions on the 
moduli spaces, may be organised into representations of
 $U(2) \semidir \bR^4$.
 Such representations 
are labelled by a magnetic label specifying the orbit of the 
magnetic charge under the  $U(2)$ action and by an  electric label
specifying the representation of the magnetic charge's  centraliser 
subgroup. In fact, the interplay between orbits and centraliser
representations is  familiar in the  theory  of induced representations
of regular semi-direct products (most famously in the representation
theory of the Poincar\'e group). 
Not all representations of  $U(2) \semidir \bR^4$
arise, however. Rather, the Dirac quantisation condition selects certain
representations and also imposes restrictions on the representations
which may be multiplied with each other. With these restrictions
we get a complete and consistent description of the dyon spectrum,
and can compute the  Clebsch-Gordan coefficients for tensor products
of dyonic states. We  emphasise   that at this point the 
full magnetic orbits play a crucial role. A number of authors
have argued that only the orbits of the magnetic weight
vectors under the Weyl group of  the exact symmetry group $H$ 
are physically relevant. Here we shall see that one cannot understand
the fusion of two charge one monopoles carrying $U(1)$ charges
into a charge two monopole carrying $U(2)$ charge without including
the magnetic orbit in the discussion.

In Sect.~11 we  
turn to a  discussion  of electric-magnetic duality or, more generally,
S-duality in $SU(3)$
Yang-Mills-Higgs theory. By  S-duality we mean the 
generalisation of 
 original electric-magnetic duality conjecture  by Montonen and Olive
\MO\  which  takes into account the $\theta$-term and the resulting
Witten effect and which is believed to be an exact duality of 
${\cal N}=4$ supersymmetric Yang-Mills theory. This version has so far
 almost exclusively been studied   in the case   where the unbroken gauge
group is abelian (see, however, \Swansea\ and our comments at 
the end of Sect.~11).   If the unbroken gauge group $H$ is non-abelian,
on the other hand,  a generalisation of the 
 abelian electric-magnetic duality conjecture was 
formulated by  Goddard, Nuyts and Olive in  \GNO.
According to the GNO  conjecture 
the magnetically charged states of the theory fall 
into representations
of the dual group $\tilde H$  of the unbroken gauge group  $H$.
The full symmetry group  of the theory would then be $H\times\tilde H$.
However, this conjecture is not compatible with the results of our 
investigation of the  semi-classical dyon spectrum.
In treating magnetic and electric properties as independent 
the representation theory of the GNO group $H\times \tilde H$ 
does not correctly account for the interplay between magnetic and 
electric charges. Our  semi-direct product group 
$H\semidir \bR^D$ ($D$ =dim$(H)$), by contrast, accurately  
captures this interplay
and moreover 
leads to Clebsch-Gordan coefficients for the combination of 
dyonic states which are consistent  with the realisation 
of the dyonic states as wavefunctions on  monopole moduli
spaces. 

We therefore study the possibility of defining an
 action of S-duality on the representations of $H\semidir \bR^D$.
In a natural implementation of this idea 
 the electric-magnetic duality
operation exchanges the two sorts of labels 
which characterise the representations of $H\semidir \bR^D$,
namely the  magnetic orbits  labels and 
the labels of  electric centraliser representations.
To test  this possibility  we set up 
a ${\cal N}=4$ supersymmetric quantisation scheme.  We show that the 
dyonic BPS states which are  the S-duality
partners of the massive  $W$-bosons in the  
minimally broken $SU(3)$ theory  can be found by  a suitable embedding of 
BPS states in $SU(2)$ Yang-Mills-Higgs theory.
Nonetheless a number of open problems remain, and these are discussed
in the final Sect.~12.

\newsec{A review of $SU(3)$ monopoles}

A monopole solution of 
 $SU(3)$  Yang-Mills-Higgs theory with coupling constant $e$
 in the Bogomol'nyi
limit is a   pair $(A_i,\Phi)$
of a SU(3) connection $A_i$ and an adjoint  Higgs field $\Phi$
on $\bR^3$
satisfying 
the Bogomol'nyi equations
\eqn\bog{\eqalign{
D_i\Phi = B_i
}} 
as well as certain boundary conditions, to be specified below.
We use the notation $\br$ for a vector in  $\bR^3$, $\partial_i$
for partial derivatives with respect to its Cartesian components,
which we also sometimes write as $(x,y,z)$, and $r=|\br|$ for its
length. In writing \bog\ we 
have also  used the usual notation 
$D_i= \partial_i + e A_i$ for  the covariant derivative
and $B_i$ for  the non-abelian magnetic field:
\eqn\magn{\eqalign{
B_i = {1\over 2} \epsilon_{ijk}\left(\partial_j A_k -\partial_k A_j
+ e [A_j, A_k]\right).
}}
To discuss solutions of this equation, in particular the 
precise boundary conditions, we need to  introduce some notation
for the Lie algebra $su(3)$ of $SU(3)$.

A Cartan subalgebra (CSA) of $su(3)$
is given by the set of diagonal, traceless $3\times 3$
matrices, and for definiteness we choose the   
 following basis $\{H_1,H_2\}$,
 normalised so that $\tr \left(H_{\mu}  H_{\nu}\right)
 = {1\over 2} \delta_{\mu \nu}$,  $\mu,\nu =1,2$.
\eqn\csa{\eqalign{
H_1 = {1\over 2} \pmatrix{1 &0&0\cr0 & -1 & 0 \cr0 & 0 & 0}
\quad\hbox{and}\quad 
H_2 = {1\over 2 \sqrt 3} \pmatrix{1  & 0 & 0 \cr 0 & 1 & 0 \cr 0 & 0 & -2}.
}}
Often we also write $\bH = (H_1,H_2)$.
We  complement the  Cartan generators  by  ladder operators
$E_{\bett}$, one for each of  the roots  $\bett=(\beta_1,\beta_2)$, 
thus obtaining
a  Cartan-Weyl  basis of $su(3)$:
\eqn\ladder{\eqalign{
 [H_{\mu}, E_{\bett}] = \beta_{\mu} E_{\bett} \quad \hbox{and} \quad
[E_{\bett}, E_{-\bett}] = 2 \bett\cd \bH.
}}
As simple roots one may take for example 
\eqn\siroots{\eqalign{
\bett_1 = (1,0) \quad  \hbox{and} \quad 
 \bett_2 =(-{1\over 2},{\sqrt3 \over 2}).
}}
All roots can be written as integer linear combinations
of these with either only positive or only negative coefficients.
It is also important for us that for  any given root $\bett$
the elements $\bett\cd\bH, E_{\bett},E_{-\bett}$ satisfy the 
commutation relations of $SU(2)$
with $E_{\pm\bett}$ playing the roles of raising and lowering
operators. In particular, we will work with the $SU(2)$ algebras
associated to the simple roots, so we write  down the associated 
generators explicitly. For
 $\bett_1$ we define
\eqn\sugen{
I_3= \bett_1\cd\bH = H_1, \quad I_+= E_{\bett_1}=
 \pmatrix{0 &1 &0\cr0 & 0  & 0 \cr0 & 0 & 0}, \quad
I_- =  E_{-\bett_1}=
 \pmatrix{0 &0 &0\cr1  & 0  & 0 \cr0 & 0 & 0},
}
and introduce $I_1= (I_+ + I_-)/2$ and $I_2 =(I_+ - I_-)/2i$, as well as 
the vector notation $\bI = (I_1,I_2,I_3)$.
For $\bett_2$ we write similarly
\eqn\sugenu{
U_3= \bett_2\cd\bH ={1\over 2}  \pmatrix{0 &0  &0\cr0 & 1  & 0 \cr0 & 0 & -1}
 \quad U_+= E_{\bett_2}=
 \pmatrix{0 &0  &0\cr0 & 0  & 1 \cr0 & 0 & 0}, \quad
U_- =  E_{-\bett_2}=
 \pmatrix{0 &0 &0\cr0  & 0  & 0 \cr0 & 1 & 0},
}
and also introduce $U_1= (U_+ + U_-)/2$ and $U_2 =(U_+ - U_-)/2i$.
Finally we define  the ``hypercharge'' operator
\eqn\ugen{
Y = {2\over \sqrt 3} H_2
}
and note that $Y,I_1,I_2,I_3$ satisfy the $u(2)$ Lie algebra
commutation relations.  Later
we will need an explicit parametrisation of the $U(2)$ subgroup
of $SU(3)$ generated by $I_1,I_2,I_3$ and $Y$, so we define Euler angles by
writing an arbitrary element $P$  in that subgroup as
\eqn\euler{
P(\chi,\alpha,\beta,\gamma) = e^{i\chi Y} e^{-i\alpha I_3} e^{-i\beta I_2}
e^{-i\gamma I_3}.
}
The  ranges for the angles are  $\chi \in [0,2\pi),
\alpha \in [0,2\pi), \beta \in[0,\pi)$ and $\gamma \in [0,4\pi)$,
supplemented by the $\bZ_2$-identification $(\chi,\gamma) \sim
(\chi +\pi, \gamma + 2\pi)$.

The boundary condition we impose on solutions  of \bog\ are
of different  types. The first sets the symmetry breaking scale:
\eqn\syca{
|\phi|^2 \rightarrow {1\over 2} v^2 \quad \hbox{for} \,\, r \rightarrow
\infty,
}
where we have used the $su(3)$ norm $|\phi|^2 =-{1\over 2}\tr \phi^2$
The second 
 stems from the wish to keep the potential energy
\eqn\enfunc{
E(A_i,\Phi) = -{1\over 2} \int d^3 x \, \tr\left( D_i\Phi^2\right) + 
\tr\left(B_i^2\right)
}
finite. To this end we impose
\eqn\bound{\eqalign{
|B_i| = |D_i\Phi| = {\cal  O}({1\over r^2}) \quad 
\quad \hbox{for large } r.
}}
Finally   we demand that 
the  Higgs field has the following form 
along the positive z-axis:
\eqn\hasy{\eqalign{
\Phi(0,0,z) = \Phi_0 - {G_0 \over 4 \pi  z} +{\cal  O}({1\over z^2}),
}}
where   $\Phi_0$ is a constant element of $su(3)$, chosen to lie
in the CSA  so that we may define $\bh$ via $\Phi_0 =i \bh \cd \bH$.
 It  follows from the Bogomol'nyi equation \bog\ 
 that  $G_0$ commutes with $\Phi_0$ \Wein, and this 
makes it possible to also demand that $G_0$ is in the CSA.
However, we will  
not impose this requirement at this stage, which  in the mathematical
literature is called framing.
 
The constant part $\Phi_0$ determines the symmetry breaking pattern:
the unbroken gauge group is the  subgroup of $SU(3)$
which commutes with  $\Phi_0$ (its centraliser).
 If $\bh$ has a non-vanishing
inner product with both the simple roots $\siroots$ the symmetry 
is broken maximally to the  $U(1)\times U(1)$ group 
generated by the  CSA. In that case 
$\Phi_0$ has three distinct eigenvalues. 
If $\bh$ is orthogonal to either of the simple roots, 
say $\bett_1$,  then 
the symmetry is broken to the $U(2)$ subgroup
of $SU(3)$ defined earlier. We say the symmetry is minimally 
broken.  In this case, which is our main 
concern in this paper,  $\Phi_0$ has a repeated eigenvalue,
and for definiteness  we shall then  assume the following form 
\eqn\vev{
\Phi_0 = i v H_2,
}
where  the vacuum expectation value $v$  sets the scale 
for the masses of all particles in the theory.

The Bogomol'nyi equation relates the coefficient  $G_0$ to
the coefficient of the long range part
 of the magnetic field so that along the positive $z$-axis
\eqn\longmag{\eqalign{
B_3(0,0,z) = {G_0  \over 4 \pi  z^2} +{\cal  O}\left({1\over z^3}\right).
}}
Thus we may call $G_0$ the vector  magnetic charge. 
According to the generalised Dirac quantisation condition \EW, \GNO, 
the vector magnetic charge has to satisfy an integrality condition.
This is easily expressed  after rotating $G_0$ into the  
CSA, thus obtaining a Lie-algebra element which we write in terms 
a two-component vector $\bg$  as  $i \bg \cd \bH$ (the vector $\bg$
is not, in general, uniquely defined, but this does not matter here).
In the case of minimal symmetry breaking this rotation
can always be effected by the action of the unbroken gauge group,
and in the case of maximal symmetry breaking $G_0$
is automatically  in the CSA by virtue of 
 the vanishing of the commutator $[\Phi_0, G_0]$.
The generalised
Dirac condition is the requirement that  
 $\bg$   lies
in the dual root lattice, which is spanned
by the vectors 
\eqn\dual{\eqalign{
\bett_1^{*} = { \bett_1 \over \bett_1 \cd \bett_1} \quad \hbox{and}
\quad \bett_2^{*} = { \bett_2 \over \bett_2 \cd \bett_2}.
}}
(with our normalisation thus $\bett_1^*=\bett_1, \bett_2^*=\bett_2$,
but we keep the notational distinction for clarity).
According to the Dirac condition
 there exist   integers $m_1$ and $m_2$ such that 
\eqn\dirac{\eqalign{
\bg  = {4\pi \over e} \left( m_1\bett^{*}_1 + m_2\bett^{*}_2 \right).
}}
We will not review the derivation of the Dirac condition
here but note that  in  the physics 
literature it is usually derived without reference
to the symmetry breaking pattern. As we will see in
the following section, however, the mathematical 
status of the two integers $m_1$ and $m_2$ depends on 
the way the gauge symmetry is broken.

\newsec{Configuration spaces and moduli spaces}

To clarify  the significance of the integers $m_1$ and $m_2$
which appear in the Dirac condition 
we define the (infinite dimensional) space ${\cal A}_{\bh}$
as the space  or pairs $(A_i, \Phi)$ which satisfy the boundary
condition \hasy\ for some fixed element $\Phi_0 =i \bh\cd \bH$  of the CSA.
This space is not acted on by the group of all static  gauge transformations
(which is the space of all maps from $\bR^3$ to $SU(3)$) but only
those whose  limit for $z \rightarrow \infty$ commutes with $\Phi_0$.
In particular it is therefore acted on by the group of framed 
gauge transformations ${\cal G}_0$, which is defined as
\eqn\framga{
{\cal G}_0 = \bigl\{
g:  \bR^3 \rightarrow SU(3)\, \big \vert \lim_{z \rightarrow \infty}
 g(0,0,z) = \hbox{id}\bigr\}.
}
We may thus define the framed configuration space ${\cal C}_{\bh}$ as the 
quotient ${\cal A}_{\bh}/{\cal G}_0 $. This space is in general not 
connected but partitioned into  disjoint sectors  whose labels are
called topological charges. The topological charges
are  elements
of the second homotopy group of the quotient of $SU(3)$ by the 
unbroken gauge group. Thus, since  $\Pi_2\bigl(SU(3)/(U(1)\times U(1))\bigr)
 = \bZ^2$
and $\Pi_2\bigl(SU(3)/U(2)\bigr) = \bZ$, the topological  charges are  pairs
of integers in the  case of maximal symmetry breaking and a   single
integer in the case of minimal symmetry breaking. 
It follows from the results of \Taubes\ that  in the former case 
 these are the two integers $m_1$ and $m_2$
appearing  in the expansion \dirac\ and in the latter case  this is the integer
coefficient $m_2$ of  the root $\bett_2$ which is not orthogonal to $\bh$.

For minimal symmetry breaking, the vector magnetic charge
$G_0$   is in general  not invariant under the action of 
 the unbroken $U(2)$ gauge group
and it is therefore not surprising  that only the invariant
component $\tr (G_0 \Phi_0)$ has a topological interpretation.
The orbit
of $G_0$ under the action of $U(2)$ 
is trivial only if $G_0$ is parallel to $\Phi_0$; otherwise it 
is a two-sphere in the Lie algebra
of $SU(3)$ which we call the magnetic orbit. 
The magnetic orbit
intersects the CSA in two
points which are related by  a Weyl reflection, and
 for any configuration $(A_i,\Phi)$ in ${\cal A}_{\bh}$ these 
two points  again have to satisfy  the Dirac
condition.   Thus, for given topological charge $m_2$
we can label a magnetic orbit by a pair of integers
$[m_1] = \{m_1, m_2-m_1\}$.
The integers $\{m_1, m_2-m_1\}$ are 
called   magnetic weights in the physics literature 
and  the pair of integers  $[m_1]$ is called  a holomorphic charge 
in the mathematics literature. 
As we mentioned earlier the  quantisation of the magnetic
weights has a more subtle mathematical origin than the 
quantisation of the topological magnetic charges.
For a precise mathematical discussion we refer the reader 
to \Murray. The important point is  that  the definition of 
holomorphic charges requires the connection at infinity
as well as the Higgs field. As a result 
it is possible for the holomorphic
charges to change in a continuous deformation of the fields which
changes the connection at infinity. Unlike topological charges,
holomorphic charges may jump along a path in the configuration space 
 ${\cal C}_{\bh}$.

Later we need an explicit parametrisation of  the magnetic orbits.
Consider the orbit  labelled
by  $([m_1], m_2)$.  Assume without loss of generality
that $m_1$ is the larger of the two numbers in $[m_1]$. Then define
$\bg$ as in \dirac\ and write a general point on the magnetic orbit as  
\eqn\gnought{
G_0 = i P \,\bg\cd \bH \,P^{-1}
}
with $P$ defined as in \euler. Computing explicitly one finds
\eqn\magorb{
G_0= {4 m_2 \pi i  \over e }
\left( {\sqrt 3\over 2} H_2 \right) + 
{4\pi i \over e} \bk\cd \bI,
}
where the vector $\bk=(k_1,k_2,k_3)$ has the  length 
$k=|\bk| =|m_1 -{m_2 \over 2}|$ and the 
direction  given 
by $(\alpha,\beta)$:
\eqn\magvec{
\hat \bk = (\sin \beta \cos \alpha, 
\sin \beta \sin \alpha, \cos \beta).
}
The magnetic orbits play a crucial role in  the remainder 
of this paper and we therefore  switch to a labelling which refers
directly to their geometry.
Formula \magorb\ shows that we may picture the magnetic orbits as 
two-spheres in $su(3)$ with  quantised radii
$k=0,{1\over 2},1,{3\over 2} , ...$ and centres on the  one-dimensional
lattice $\{{4  \pi K i \over  e }\left( {\sqrt 3\over 2} H_2 \right) 
|K \in \bZ\}$. Thus it is natural to introduce  labels
$(K,k)$, related to 
 $([m_1],m_2) $  via 
\eqn\orlab{
K=m_2 \quad\hbox{and} \quad  k=|m_1 -{m_2 \over 2}|.
}
For the rest of the paper  we will call the vector $\bk$ 
the non-abelian magnetic charge.

To summarise: in the case of maximal
symmetry breaking the configuration space ${\cal C}_{\bh}$
is partitioned into disjoint topological sectors labelled by the 
pair of integers $(m_1,m_2)$ but in the case of 
minimal symmetry breaking the disjoint components of   ${\cal C}_{\bh}$
are only labelled by one  integer $K$. There exists a finer subdivision
according  to the magnetic weight but this is not of a topological
nature. Instead it is an example of  a stratification of a space,
with different strata labelled  by the magnetic weights.
We will describe this concretely at  the level of
solutions of the Bogomol'nyi equations, or more precisely of 
certain sets of solutions, called  moduli spaces.

Taubes was the first to establish a rigorous existence 
theorem for monopole solutions in Yang-Mills-Higgs theory
for a  general gauge group   and general symmetry breaking
pattern. In \Taubess\ he proved  the 
existence of solutions in each of the topological components of the 
framed
configuration space.
Applied to our situation this shows that for maximal symmetry 
breaking monopoles exist for all positive integers $m_1$ and $m_2$, but 
for minimal breaking Taubes' result only implies the existence 
for given positive  topological charge $m_2$
(the positivity condition results from our choice of sign
in the Bogomol'nyi equations; if we had studied the equation
with the opposite sign we would obtain anti-monopole solutions
with negative topological charges). Taubes' results say nothing 
about the existence of  monopoles with  given   magnetic weight.
On the other hand, a number of explicit solutions have been
known for some time. The search for  monopole solutions in Yang-Mills
theory (with or without Higgs field) 
for  general gauge groups has a long history, 
see \ansat\ for   the case of $SU(3)$. 
It was also noticed early on
 that  certain solutions  of the Bogomol'nyi equations  of the 
critically coupled  Yang-Mills-Higgs  theory
can be  obtained  
by deformations of  embedded $SU(2)$ monopole solutions, a possibility 
first pointed out in \Sander\  and discussed further by Weinberg \Wein\ and 
Ward  \Richh. In the latter paper it is shown that in the minimally broken
$SU(3)$ theory   solutions of arbitrary topological charge
$m_2$  can be obtained by
embeddings of $SU(2)$ monopoles  but that the magnetic weight is 
necessarily $[m_1]=[0]$; we will study these solutions
in detail below. Bais and Weldon found the first solution which
cannot be constructed by embedding an $SU(2)$ solution \BW. This solution
has topological charge $m_2 =2$  and magnetic weight $[m_1]=[1]$;
it is spherically symmetric and is part of a  
 six parameter  family of axisymmetric solutions later found
by Ward \Rich, which we will discuss in detail  later in this paper.

Monopole solutions of the same topological characteristic
are conveniently grouped together in moduli spaces.
In the case of maximal symmetry breaking there is a canonical
way of defining the moduli spaces. For given $\Phi_0$ in the 
expansion \hasy\ we fix the topological labels $(m_1,m_2)$
and consider the set of all solutions of the Bogomol'nyi equation
in the corresponding  component of the configuration space. In 
symbols:
\eqn\abmod{
M^{\footnotefont{max}}_{m_1,m_2}
= \bigl\{ (A_i,\Phi)\in {\cal A}_{\bh}\,\big \vert  D_i\Phi = B_i,\,\,
\bg  = {4\pi \over e} \left( m_1\bett^{*}_1 + m_2\bett^{*}_2 \right)
\bigr\}\,\, /
{\cal G}_0.
}
It follows from Taubes' existence theorem that there is a 
non-empty moduli space of monopole
solutions for  each point in the dual 
root lattice  shown in Fig.~1.a with positive coordinates $(m_1,m_2)$.
Weinberg \Wein\ long ago 
 counted  how many parameters' worth of solutions there
are for each point in the dual root lattice.
  Translated into our language, this determines the dimension
of the moduli spaces. Observing carefully  the
slight differences in conventions, Weinberg's results 
translate into
\eqn\dimcount{
\hbox{dim}\,M_{m_1,m_2}^{\footnotefont{max}} = 4(m_1 + m_2).
} 
As also pointed out by Weinberg, this dimension formula 
suggests that there exist multi-monopole solutions in
this theory which are made up of $m_1$ well-separated $SU(2)$  monopoles
embedded along the root $\bett_1$ and $m_2$
well-separated  $SU(2)$ monopoles
embedded along the root $\bett_2$.

In the case of minimal  symmetry breaking  
 only the component of $G_0$ parallel to $\Phi_0$, labelled by
 the integer $m_2$, 
has a  topological significance. Thus one may define the 
corresponding  moduli spaces as 
\eqn\nabmod{
M_{m_2}= \bigl\{ (A_i,\Phi)\in {\cal A}_{\bh}\, \big \vert D_i\Phi = B_i,\,
- 2 \tr(G_0 \Phi_0) = { 4 \pi m_2 i \over   e} \bett_2\cd \bh  \bigr \}\,\, /
{\cal G}_0.
}
However, as we learnt earlier we may classify finite-energy 
configurations   further
according to the magnetic weight. In the mathematical
literature it is customary to denote 
 the set of all monopoles 
in  $M_{m_2}$ with magnetic weight $[m_1]$ by $M_{[m_1],m_2}$.
Then the moduli spaces are labelled by the same labels as the 
magnetic orbits, and at the risk of confusing mathematicians 
who may read this paper we will use our preferred orbit labels
$(K,k)$ \orlab\ also for the moduli spaces, so we write $M_{K,k}$
instead of $M_{[m_1],m_2}$.
As anticipated earlier  the spaces $M_{K,k}$  are components of 
the {\it   connected}  space $M_{K}$.  In mathematical
terminology one says that the space $M_{K}$ is stratified 
with strata $M_{K,k}$. 
It was first pointed out by Bowman \Bowman\ that counting 
and interpreting the parameters of monopoles in the 
case of less than maximal symmetry breaking is considerably
more complicated than in the case of maximal symmetry breaking.
Murray was the first systematically to compute the dimension of 
the moduli spaces  of solutions in \Murray.
The situation is summarised in Fig.~1.b.

We can still think of  each point of the dual root lattice
as representing a (possibly empty) moduli space
of solutions, but 
we need to keep in mind that  points related by reflection at the 
vertical axis (which maps $m_1$ to $m_2-m_1$) represent the same
moduli space.
The results of \Murray\  imply that there only exist monopole solutions 
if the size of the magnetic orbit is 
less than or equal to  the (positive) topological charge $K$
which means that $k=0,1, ..., {K\over 2} $ if $K$ is even and 
$k={1\over 2}, {3\over 2}, ..., {K\over 2}$ if $K$ is odd.
 Thus solutions only occur
inside the cone (including the edge) drawn in Fig. 1.b, 
where we also give the dimension
which Murray computed for  each of the non-empty moduli spaces. 
Interpreting 
those dimensions physically  is one of the objectives of this paper. 
To that end, however, we need more explicit descriptions of 
the moduli spaces. We are particularly interested in the moduli
spaces on the edge of the cone and, for even $K$, in the 
centre of the cone. For given $K$ these are the strata of
respectively smallest  and  largest dimensions. Hence  they are
referred to respectively as the small and the large strata.

Before  we give  more explicit descriptions of the moduli 
spaces we note that the moduli spaces we have defined
are {\it a priori} only defined as sets. While there may be
various mathematically natural ways to give these sets the 
structure of a manifold, the physically relevant structures are
induced from the underlying field theory. In particular one
would like to induce  the structure of a differentiable manifold from
the field theory configuration space 
and  define a Riemannian metric   from the field theory 
kinetic energy functional.  This  works well for example 
in the case of $SU(2)$ monopole moduli spaces \athi\ but 
is known to be problematic for solitons in the $\bCP^1$-model \wardcp.
In the discussion of the moduli spaces $M_{K,k}$ we will
also encounter pathologies which are in some sense worse than
in the case of $\bCP^1$ lumps. However, in order to describe 
these pathologies explicitly we recall here the formal definition
and general form of the metric on the moduli spaces. 
Consider therefore some generic moduli space  $M$  of monopoles and 
suppose we have  introduced (local)
 coordinates $X=(X_1,  ..., X_{\hbox{\footnotefont dim}\, M})$
 with components 
 $X_{\alpha}$, $\alpha = 1, ..., \hbox{dim}\, M$
on $M$. Then  a monopole configuration in $M$ may be written
as $(A_i,\Phi)(X;\br)$, exhibiting explicitly the dependence
on both the collective coordinate $X$ and the spatial coordinate $\br$.
The metric $g_{\alpha\beta}(X)$ is defined via the 
$L^2$ norm of the infinitesimal variations $(\delta_{\alpha}A_i,\delta_{\alpha}
\Phi)$ which are required to satisfy the linearised Bogomol'nyi
equations and, crucially, Gauss' law.
In the gauge $A_0 = 0$  this reads
\eqn\gauss{
D_i\delta_{\alpha}A_i + [\Phi,\delta_{\alpha}\Phi] = 0.
}
Then the metric can in 
principle be computed via
\eqn\metric{
g_{\alpha\beta}(X) = -{1\over 2}\int  d^3 x \,\tr\left( \delta_{\alpha} A_i
\delta_{\beta}A_i\right) +
 \tr\left(\delta_{\alpha}\Phi \delta_{\beta}\Phi \right).
}
Note that by construction the Euclidean group $E_3$ of translations
and rotations in $\bR^3$ and the unbroken gauge group 
act on the moduli spaces  we have defined and that the above
expression is formally invariant under those group actions.
Thus we expect the Euclidean group and the unbroken gauge group
to act isometrically on  the moduli spaces
in all cases where the above metric is   well-defined.

\newsec{Embedding SU(2)  monopoles}

Before we enter the general discussion of the
structure of the moduli spaces  we
note that a large family of $SU(3)$  monopole  solutions can be obtained 
by simply embedding $SU(2)$ monopoles. This family will
be particularly important for us. The method is simple: 
take a simple root which has a positive inner product with
$\Phi_0$ and embed the monopole in the associated $SU(2)$
Lie algebra, adding a constant Lie algebra element to satisfy
the boundary condition \hasy. In discussing the explicit form
of  solutions we will take the value of  $v$ \vev\ to be 
${1\over2\sqrt 3}$
in order to make contact with other explicit solutions discussed in \And.
Thus, taking the simple root $\bett_2$ for definiteness
and referring to the definitions \sugenu\ of the generators
$U_l$, $l=1,2,3$,   we define
\eqn\embed{\eqalign{
\Phi^u& = \sum_{l=1}^3 \phi_l U_l + \hbox{diag} 
(1,-{\textstyle{1 \over 2}},-{\textstyle{1 \over 2}})
\cr
A_i^u & = \sum_{l=1}^3 a_{il} U_l,
}}
where $(a_i, \phi)$ is a $SU(2)$  monopole of charge 
$K$ scaled so that  its Higgs field
tends to diag$({3\over 2}, -{3\over 2})$ along the positive $z$-axis. 
 The Higgs field
of the embedded  solution then has the following  expansion 
along the positive $z$-axis
\eqn\emexp{
\Phi^u = \Phi_0 - {K \over   e  z} U_3 + {\cal O}
\left(1\over z^2\right),
}
showing that  the embedded solution is an element of the space
$M_{K,K/2}$.

Note, however, that this embedding is not unique. We obtain
an equally valid solution after  acting with the unbroken
symmetry group $U(2)$. In terms of the explicitly parametrised
element $P$ in \euler\ we define
\eqn\rotemb{\eqalign{
\Phi_P &=P \Phi^u P^{-1} \cr
A_i^P &= P A_i^u P^{-1}.
}}
What is the orbit  under the action of $P$?
First note that 
\eqn\commus{
[Y, U_{\pm}] = \pm U_{\pm} \quad \hbox{and} \quad 
[I_3, U_{\pm}] = \mp {1\over 2} U_{\mp}
}
and hence that the configuration \embed\
  is invariant under the $U(1)$ subgroup 
generated by  $Y+ 2 I_3$. It follows that the orbit 
of a given embedded configuration
$(A_i^u,\Phi^u)$ under the $U(2)$  action is the quotient of $U(2)$
by that $U(1)$ group; this is   a three-sphere  which we denote by
 $S_P^3$. This three-sphere 
can be coordinatised in a  physically meaningful way 
using the Euler angles $(\alpha,\beta,\gamma)$ as defined in \euler\
(the angle $\chi$ is redundant).
Alternatively we can think of elements of  $S^3_P$ as $SU(2)$ matrices
$Q$, given by 
\eqn\qmat{
Q(\alpha,\beta,\gamma)  = e^{-{i\over 2} \alpha \tau_3}
e^{-{i\over 2} \beta \tau_2}
e^{-{i\over 2} \gamma \tau_3}
}
 This three-sphere is Hopf-fibred
over the   magnetic  two-sphere, defined in \gnought\
and we can now write the corresponding Hopf map
$\pi_{\footnotefont{Hopf}}^k$, which depends on the 
magnitude of the non-abelian magnetic charge:
\eqn\hopfi{
\pi_{\footnotefont{Hopf}}^k:Q\in S^3_P \rightarrow \bk \in S^2,
}
with $\bk$ defined as in \magvec.
The angle  $\gamma$ parametrises the 
`body-fixed' $U(1)$ rotations about the 
vector $\bk$ \magorb\ 
which only change the monopole's short-range fields. 
The role of this circle  is well-understood in
the context of $SU(2)$ monopoles: motion around it gives the monopole
electric charge. Thus we will call the circle parametrised by $\gamma$
the electric circle. Then we can sum up the preceding discussion by
saying that  the three-sphere $S^3_P$ is  Hopf-fibred over the 
magnetic orbit $S^2$ with fibre the electric circle.

The embedding procedure just described can be carried 
out for $SU(2)$ monopoles of arbitrary charge. The moduli
space of the latter  is well-understood and for magnetic charge $K$
it  has the form
\eqn\sutwo{
M^{\footnotefont SU2}_K = \bR^3 \times {S^1 \times M^0_K \over \bZ_K}
}
where $\bR^3$ coordinatises the centre-of-mass of the charge
$k$ monopoles, the $S^1$-factor is the electric circle introduced 
at the end of the previous paragraph and 
$M^0_K$ is the $K$-fold cover of the moduli space
of centred (both in $\bR^2$ and $S^1$) $SU(2)$ monopoles of charge $K$.
The space $M^{\footnotefont SU2}_K$ has dimension $4K$.
 As mentioned earlier, embedding $SU(2)$
monopoles gives rise to    $SU(3)$ monopoles of the same 
topological magnetic  charge and with  magnetic orbit radius $K/2$. 
In other words embeddings of $SU(2)$ give families
of $SU(3)$  monopoles which are elements of 
the small strata of the moduli space of $SU(3)$ monopoles. 
In fact it is easy to see from
the Nahm data  (see e.g. \Bowman) that all $SU(3)$  monopoles
in the small  strata can be obtained via embeddings.
Thus putting  together  our explicit embedding prescription
with the formula \sutwo\ 
we  deduce that  the small  strata $M_{K,K/2}$ are fibred over
the magnetic orbit   with  the spaces $M^{\footnotefont SU2}_K$
as fibres. We thus  have the bijection
\eqn\emod{
M_{K,K/2} \leftrightarrow \bR^3 \times  {S^3_P \times M^0_K \over \bZ_K}.
}
Here
$\bZ_K$ acts on $S_P^3$ by moving a fraction $2\pi/K$ round the fibre
of this fibration (the electric circle). In the fibration
\eqn\fibre{
\matrix{
M^{\footnotefont SU2}_K & \longrightarrow & M_{K,K/2} \cr
                         &                & \mapd \cr 
                        &                  & S^2 \cr }
}
the projection map $\pi^k$ is the forgetful map on  $ \bR^3$ and $M^0_K$
and the Hopf map \hopfi\ on $S_P^3$. For later use we also introduce
the notation $M_{K,\bk}$ for  the fibre $(\pi^k)^{-1}(\bk)$.
Note in particular 
that the number of independent parameters in the 
 spaces $M_{K,K/2}$ is   $4K + 2$, 
thus agreeing 
with the dimension found for the small  strata  by Murray.

We have not yet said  anything about the differentiable  and metric
structure which $M_{K,K/2}$
inherits from the field theory kinetic energy functional via
\metric. By inspection one checks  that the moduli space
metric \metric\ is well-defined on  the fibres of the fibration
\fibre\ and that, with that metric, the fibres  
are (up to an overall scaling  factor of  $1/3$ required to agree
with the conventions of \athi)
 isometric to the $SU(2)$ monopole moduli spaces.
However, the mathematical 
structure of the magnetic orbit harbours a 
 number of 
surprises and subtleties, most  of them related to 
physical  observations made some time ago and all of them
to do with the action of the unbroken gauge group  $U(2)$.
Since this group action is of central importance 
for our investigation we  discuss it in a separate section.

\newsec{`Global Colour' revisited}

It is a standard lore in the theory of topological defects
that if a defect breaks a symmetry of the underlying theory
the broken symmetry  generators can be used to define
collective coordinates for the defect. However
it was realised long ago by a number of authors  that 
this is problematic  in gauge theories
when  the gauge symmetry gets broken to a non-abelian group.
Historically this discussion was mostly conducted in
 the context of grand unified theories with gauge group $SU(5)$
broken to $SU(3)_{\footnotefont  colour}\times 
U(2)_{\footnotefont  electroweak}$
and in that context the question of defining and dynamically
exciting the collective coordinates associated to the unbroken
gauge group was put succinctly by Abouelsaood : ``Are there
chromodyons ?'' \Abouu.

The answer was  given partly 
by Abouelsaood himself and complemented by the important observation
of Nelson and Manohar 
\NM\ (see also \BMMNSZ\ ) that ``Global colour is not always
defined''. Briefly, and applied to our situation, the latter 
authors noted that if one writes down the Higgs field on 
the two-sphere at spatial infinity in  any regular  gauge
and attempts to define generators of a $U(2)$ algebra which
commute with the Higgs field everywhere and vary smoothly
over the two-sphere one will only succeed if the topological
 magnetic charge of the configuration is even. For odd 
topological charge
there is a topological obstruction 
similar to the one preventing
the existence of a smooth non-vanishing vector fields on a two-sphere;
in that case it is only possible to extend  a maximal torus of $U(2)$
over the entire two-sphere at infinity.
 The result of Nelson and Manohar \NM\ imply that if one 
insists on defining  a $U(2)$ action at a fixed point (say
$z=+\infty$) in the case
of  odd magnetic charge  then every extension of it
over the entire two-sphere at spatial infinity will  necessarily 
change not only the $1/r$ term in the asymptotic expansion of 
the Higgs field  but even the $r^0$ part 
 somewhere on the two-sphere at spatial infinity.

However, even if one allows  the $U(2)$ action to change 
the Higgs field (by a gauge transformation)  on the two-sphere 
at spatial infinity
there are  problems with the collective coordinates produced 
by the generators which do not commute with 
the vector magnetic charge $G_0$. 
These were pointed out by Abouelsaood \Abouu\ who showed 
that infinitesimal deformations produced by such generators do not
satisfy  the constraint imposed by  Gauss' law \gauss.
In the following we will call collective coordinates whose
infinitesimal variation produces zero-modes which satisfy
Gauss' law and which have a  finite $L^2$-norm (so that the 
corresponding components of the metric \metric\ are finite)
dynamically relevant.
Abouelsaood's result is thus  that collective coordinates produced
by generators of the unbroken gauge group which do no commute
with the vector magnetic charge are not dynamically relevant.

How do these observations rhyme with  our description
of monopole moduli spaces in the two preceding sections?
Returning to Fig.~1.b we first note that all moduli spaces
on the vertical axis  are of the form $M_{K,0}$, 
with   the topological magnetic charge $K$  necessarily even.
Moreover   the vector magnetic  
charge $G_0$ is parallel
to the  constant part of the Higgs field $\Phi_0$. Thus $G_0$
is by definition invariant under the action of all generators
of the unbroken gauge group and we expect no problems in 
defining the action of the unbroken gauge group  on these spaces.
For all the other moduli spaces, however,  the magnetic charge  
$G_0$ obstructs  the action of the unbroken gauge group $U(2)$.
 It then follows from the results 
described in the previous paragraph that only
the centraliser of $G_0$ in 
$U(2)$ can have a dynamically relevant action on these spaces.

The unbroken gauge group $U(2)$ explicitly entered 
our description of 
the small strata $M_{K,K/2}$
and it is therefore not difficult to isolate the 
physically problematic coordinates in those spaces.
These are by definition coordinates associated with
the generators of the unbroken gauge group which do not commute
with $G_0$ and 
are therefore precisely the coordinates on the 
magnetic orbit, i.e.  on  the 
base space of  the fibration of the  spaces $M_{K,K/2}$ 
in \fibre. Motion on the fibre  is physically unproblematic
but  motion orthogonal to the fibres is physically not permitted.
The mathematical reason for this is that the magnetic orbit 
inherits neither a differentiable nor a metric structure from
 the field theory. On the other hand, thinking of the magnetic
orbit as a two-sphere in the Lie algebra $su(3)$ it is natural,
in view of our remarks after \magvec, 
to induce mathematical structure from this embedding. 
In the quantum theory, to be discussed in Sect.~10, we will
indeed require some mathematical structure on the magnetic orbit,
namely an integration measure (which does not presuppose the 
existence of a  metric structure). Thinking of the magnetic orbits
as round two-spheres  of radius $k$ 
(this was anticipated in defining the Hopf map
$\pi_{\footnotefont{Hopf}}^k$   in \hopfi)
we  thus define the integration
measure $k^2 \sin\beta \, d\beta \wedge d\alpha$ on them.
Although we have only been able to isolate the magnetic
orbit explicitly as part of the moduli space in the smallest
strata  we  expect on physical grounds  all strata $M_{K,k}$
with $k >  0$ to be fibred over two-spheres parametrising the 
non-abelian magnetic charges. 
This conjecture  does not appear to have been considered 
 in the mathematics literature and we are not able to write down  the 
projection maps for these fibrations explicitly. 
Nontheless 
we shall assume that projection maps exist for all $M_{K,k}$
provided $k>0$ (in this paper we will only use them for $k=K/2$)
and think of the base spaces of these fibrations as round two-spheres
with  the integration measure given above.

The observations of Abouelsaood, Nelson and Manohar have  lead 
most authors discussing $SU(3)$ monopoles 
in the literature to  discard the
 magnetic orbit altogether, see \alvar\
for a recent example.
 For us there are two reasons for
keeping the magnetic orbit in the discussion, and for
equipping it with the measure given above. 
The first is that  from
a certain  mathematical point of view, to be described in
Sect.~7, it is very natural to include the magnetic  orbit in the 
moduli spaces. The second and more important reason is that 
the magnetic orbit plays a crucial  role in the  full understanding of  
 of (classical and quantised) dyonic excitations and of 
the behaviour of several interacting monopoles.
In  particular we will see that is impossible to understand 
how two quantum states of  topological charge one monopoles 
combine to a  quantum state of a  topological charge two
monopole without taking the magnetic orbit into account.
 Much of the 
remainder of this paper is devoted to explaining this point,
but as a first step we use the next, short section   to exhibit
some of the elementary questions  which arise when studying classical
interacting monopoles with non-abelian magnetic charge.

\newsec{Counting monopoles and their moduli}

To begin,  focus    on the moduli
space $M_{1,1/2} = \bR^3 \times S^3_P $. Physically this space summarises
the degrees of freedom of a single monopole in minimally broken
$SU(3)$ Yang-Mills-Higgs theory and is therefore the basic building
block for any understanding of monopole physics in that theory.
We have seen that of the monopole's six collective coordinates
only four are dynamical: three coordinates for the monopole's position
and one for the  electric circle. Thus, dynamically  a single
$SU(3)$  monopole has the same degrees of freedom as an $SU(2)$
monopole, but in addition it has a non-dynamical label, namely a 
point on a two-sphere which specifies the direction of the
vector  magnetic charge  in the Lie algebra of $SU(3)$.
The crucial and interesting point is, however, that this magnetic
direction and the electric degree of freedom are not independent:
the electric circle is generated by the centraliser group of 
the magnetic charge. Monopoles with different magnetic directions
therefore carry charge with respect to different $U(1)$ groups.

Now consider combining two monopoles. This can be done 
in two distinct ways, corresponding to the two strata of the 
moduli space $M_2$ of monopoles of topological charge two. To obtain
a configuration in the small stratum $M_{2,1}$ the vector magnetic 
charges of the individual monopoles have to be parallel but  to 
obtain a configuration in the large stratum  $M_{2,0}$
the vector magnetic charges should cancel and thus be anti-parallel.
More generally the restriction that the radius $k$ of the magnetic orbit
is less than or equal to half the topological magnetic charge $K$, 
pictorially
expressed in  the cone structure of Fig.~1.b, means that at least
in principle it is possible to interpret configurations in $M_K$
as being made up  of $K$ single monopoles. In particular we know already
from our discussion of the  small 
stratum $M_{K,K/2}$ that it indeed 
 contains  configurations made up of $K$
monopoles with all their vector magnetic charges aligned.
This raises the question of whether all strata have an asymptotic region
where the moduli (and in particular their number) can 
 be interpreted in terms of the moduli of  individual monopoles.
How can this be done? As a general principle we shall assume that 
in any set  of well-separated  monopoles which satisfies
the Dirac quantisation condition any subset   must also satisfy
that condition. In particular, therefore, any pair must satisfy
the Dirac condition and thus the monopoles' vector magnetic
charges must be pairwise parallel or anti-parallel. It follows
that in a collection of $K$ monopoles the magnetic vectors
of all them necessarily  lie along one line, and 
only the individual positions and electric phases can be chosen
independently. One would then expect the dimension of 
all the strata of the moduli space $M_K$ to  be $4K+2$. In 
fact this formula is only valid for the smallest strata, where we know the 
above picture to be   correct. Amongst the other strata
the strata of smallest magnetic  weight for given topological
 magnetic charge $K$ hint at  a completely different interpretation.
There the  dimension is $6K$, suggesting the physically puzzling 
 interpretation of 
\Murray\ that in these moduli spaces
the individual monopoles' vector magnetic charges have somehow
escaped the constraint imposed by the Dirac condition and have become
independent and  dynamical. 

The next sections are  devoted to a detailed investigation
of these  and other questions in the case of the 
 12-dimensional moduli space $M_{2,0}$. This  is the 
simplest moduli space which cannot be understood via embeddings and 
as a result we need to consider mathematically more sophisticated
approaches.

\newsec{Monopoles and rational maps}

The identification of monopole moduli spaces with spaces
of rational maps from $\bCP^1$
into certain flag manifolds 
 goes back to conjectures of Atiyah and 
Murray \atmu\ and was first proved in the $SU(2)$ case by Donaldson
\donaldson. Since then generalisations of Donaldson's result to
general gauge groups and symmetry breaking have been proved,
see e.g. \Jarvis\ for recent results and further references.
The results relevant for us are mostly contained in the papers
\Murray\ and \Anddd. It is explained in \Murray\ that 
the rational maps describing $SU(3)$ monopoles with minimal
 symmetry breaking to $U(2)$ are based rational 
maps from $\bCP^1$ to  $\bCP^2$. Such maps are
topologically classified by their degree, and this equals
the topological charge of the associated monopole.
Concretely the condition that the map is based 
means that the point at infinity is sent to zero,
and we may write such maps  as functions
of one complex variable  $\zeta \in \bC$ taking values 
 in ${ \bC}^2 \cup {\infty}$. Then a rational
map of degree $K$ has the form 
\eqn\rats{\eqalign{
R(z)= \left({p_1(\zeta)\over q(\zeta)}, {p_2(\zeta) \over q(\zeta)}\right),
}}
where $q$ is a polynomial of degree $K$ whose leading
coefficient is $1$ and $p_1$ and $p_2$ are polynomials
of degree less than $K$. Writing  Rat$_K$ for the space
of all based rational maps from $\bCP^1$ to  $\bCP^2$
of degree $K$ one of the results of \Murray\ is that 
there is a one-to-one correspondence between Rat$_K$
and the moduli space $M_K$ \nabmod\ of monopoles of topological charge $K$ 
in minimally broken $SU(3)$ Yang-Mills-Higgs theory.
 There is also a stratification
of Rat$_K$ which corresponds to that of the monopole moduli
spaces, thus encoding the monopoles' magnetic weights
in the rational map.  A general account   of 
how this encoding is done can be found  in \Murray; 
in the special cases we are concerned with  we will be able 
to identify the magnetic weight explicitly  without reference
to the general theory. One basic tool for understanding
the correspondence between rational maps and monopoles
is the action of the symmetry group on the moduli spaces. In our
case this is the direct product of the Euclidean group 
of translations and rotations in $\bf R^3$
and the unbroken gauge group $U(2)$. This group acts naturally
on the moduli space $M_K$ and hence  it has  an action
on Rat$_K$ as well.

The $U(2)$ action on rational maps 
is easy to write down. Parametrising a $U(2)$ matrix explicitly as 
$e^{i\chi} Q$, where  the $SU(2)$-matrix $Q$ is  parametrised as in \qmat\
and the angles $(\chi,\alpha,\beta,\gamma)$ satisfy the $\bZ_2$ condition
specified after \euler,  
the $U(2)$ action on the rational map \rats\ is 
\eqn\utwoac{
\left( \matrix{ p_1 \cr p_2} \right) \mapsto e^{i\chi} Q
 \left( \matrix{ p_1 \cr p_2} \right).
}
The construction of rational maps from monopoles requires the choice
of a preferred direction in $\bR^3$  and hence breaks the 
symmetry of Euclidean space. We follow the conventions of \Anddd\
where the preferred direction is the $x$-direction. Then 
a translation $(x,y,z) \in \bR^3$ acts on a rational map as
\eqn\transact{
R(\zeta ) \mapsto e^{3 x} R(\zeta-{i\over 2}(y+iz)).
}
The spatial  rotation group $SO(3)$ also acts on the rational maps
but this action does not concern us here (in fact only the action
of the $SO(2)$ subgroup of rotations about the $x$-axis is known
explicitly).

Consider now the simplest case of a monopole of charge one.
The associated rational map has the general form
\eqn\degone{\eqalign{
R_1(\zeta) = ({\mu_1 \over \zeta-\epsilon},{\mu_2 \over \zeta-\epsilon}).
}}
To understand the interpretation 
of the complex numbers $\mu_1,\mu_2$ and $\epsilon=\epsilon_1+i\epsilon_2$
note that his map  is  obtained from  the standard map
$(0,1/\zeta)$ by a combined translation \transact\ and 
a $U(2)$ action if we identify
\eqn\mopopos{
(x,y,z) =
{1\over 2}({1\over 3} \ln(|\mu_1|^2 + |\mu_2|^2), -\epsilon_2,\epsilon_1)
}
and 
\eqn\mopor{
{1 \over \sqrt{(|\mu_1|^2 + |\mu_2|^2)}}
 \left( \matrix{ \mu_1\cr \mu_2} \right)
=\left( \matrix{ - e^{i(\chi+ {1\over 2}\gamma- {1\over 2}\alpha)}\sin\beta \cr
 e^{i(\chi +{1\over 2}\gamma+ {1\over 2}\alpha)} \cos\beta }  \right).
}
Note in particular that the right hand side depends on 
$(\chi,\gamma)$ only in the combination $(\chi + {1\over 2} \gamma)$
and that we can extract the polar coordinates $(\alpha,\beta)$
 for the direction
of the non-abelian magnetic charge \magvec:
\eqn\hopf{
e^{-i\alpha}\tan\beta  = {\mu_1 \over \mu_2}.
}
We thus have the following
interpretation of the parameters $\mu_1,\mu_2$ and $\epsilon$
in terms of monopole moduli.
The monopole's position in the $yz$-plane is given
by the complex number $\epsilon$ 
and the $x$-coordinate is determined by the length
of the $\bC^2$ vector $(\mu_1,\mu_2)^t$.
The corresponding  unit length vector
in  $S^3$ determines the direction of the non-abelian magnetic 
charge vector  via  the Hopf projection
\eqn\hopf{
\left(\matrix{\mu_1 \cr \mu_2} \right) \rightarrow {\mu_1 \over \mu_2}
}
and the fibre of that projection is the electric circle, parametrised
by $\gamma/2 + \chi$. Again we discard the redundant angle $\chi$.

The rational maps describing monopoles of charge two have the
general form
\eqn\degtwo{\eqalign{
R_2(\zeta) = ({a + b\zeta  \over \zeta^2 + f \zeta -\epsilon^2},{c + d \zeta 
 \over \zeta^2 + f \zeta -\epsilon^2}).
}}  
The set of all such maps form a 12-dimensional manifold, parametrised
by the complex numbers $a,b,c,d,\epsilon,f$. We define the strata
of this manifold with the aid of the matrix
\eqn\mattt{\eqalign{
M = \left( \matrix{a & b \cr c & d} \right).
}}
The small stratum of Rat$_2$   is  defined by the condition
that for all its elements  the determinant of  $M$ vanishes.
Thus the small stratum is 10-dimensional and Murray showed \Murray\
that there is a one-to-one correspondence between it and
the small stratum 
$M_{2,1}$ of  the charge two $SU(3)$  monopole 
moduli space.   Geometrically the condition det$M$=0 means  
that the range of the corresponding rational 
map lies entirely inside some
${\bCP}^1 \subset {\bCP}^2$. Since $\bCP^2$ is fibred over
$\bCP^1$ with fibre $\bCP^1$ one deduces that  the small stratum 
of Rat$_2$ is also fibred 
over $\bCP^1$ with each fibre diffeomorphic to  the set of rational
maps $\bCP^1 \rightarrow \bCP^1$ of degree two.
Since the latter set is, by Donaldson's theorem, isomorphic
to the moduli spaces of charge two $SU(2)$ monopoles
we recover the structure \fibre.

Note that 
the small stratum of Rat$_2$ naturally has the structure of a 
complex differentiable manifold whereas
 in the fibration \fibre\  of the monopole moduli space 
the fibres, but not the base space, inherit 
a differentiable structure from the field theory. It follows that the
bijection between the small stratum of rational maps and the small
stratum of the monopole moduli space is
not a diffeomorphism.

The large stratum is defined as the set of maps in Rat$_2$
for which the determinant of $M$ does not vanish.
It is twelve dimensional  and was shown by Dancer \Anddd\
to be  diffeomorphic to the big stratum
 $M_{2,0}$  of the charge two 
monopole moduli space.   Note that the rational map description
of the moduli spaces greatly clarifies the relation between the 
strata. We can now see explicitly that the strata are part of 
the connected set Rat$_2$ and that in a precise sense the small
stratum is the  boundary of the large stratum.

Concentrating now on the large stratum $M_{2,0}$ 
we would again like to find a physical interpretation of the 
parameters occurring in \degtwo.
By acting on a generic rational map of the form \degtwo\
   with a suitable translation in the $yz$-plane we can set $f=0$.
Now consider those maps for which $|\epsilon|$ is large.
Then we may  write the map \degtwo\ in terms of partial fractions as
\eqn\ratfar{\eqalign{
\tilde R_2(\zeta) = \left({\mu_1 \over\zeta-\epsilon} + 
{\nu_1 \over \zeta +\epsilon},
 {\mu_2 \over
\zeta-\epsilon} +{\nu_2 \over \zeta+\epsilon}\right),
}}
with the parameters $\mu_1,\mu_2, \nu_1$ and $\nu_2$ 
related to $a,b,c,d$ and $\epsilon$ via
\eqn\aallbd{\eqalign{
\left(\matrix{\mu_1 \cr \mu_2} \right)=
 {1\over 2}\left(\matrix{  {a\over\epsilon } + b \cr
{c\over \epsilon} +d} \right) \quad \hbox{and} \quad
\left(\matrix{\nu_1 \cr \nu_2} \right)= {1\over 2}\left(
\matrix{-{a\over \epsilon} + b \cr
-{c\over \epsilon} +d} \right).
}}
The map \ratfar\ is clearly the sum of two degree one maps,
and it is tempting to interpret it as describing two monopoles located at
\eqn\twomo{\eqalign{
{1\over 2}({1\over 3}\ln  2 |\epsilon| 
(|\mu_1|^2 + |\mu_2|^2), -\epsilon_2,\epsilon_1)
\quad \hbox{and} \quad
{1\over 2}({1\over 3} \ln  2 |\epsilon|
 ( |\nu_1|^2 + |\nu_2|^2),\epsilon_2, -\epsilon_1)
}}
with internal  orientation given by the vectors $(\mu_1,\mu_2)$
and $(\nu_1, \nu_2)$ normalised to lie on the unit sphere $S^3$
in ${\bC}^2$.
In particular this suggests that 
the individual monopoles' non-abelian magnetic charges have
directions $(\alpha_1,\beta_1)$ and $(\alpha_2,\beta_2)$  given  by  
\eqn\magor{\eqalign{
 e^{-i\alpha_1}\tan \beta_1 = {\mu_1 \over \mu_2}  \quad \hbox{and}
 \quad  e^{-i\alpha_2}\tan \beta_2 = {\nu_1 \over \nu_2}.
}}
What is remarkable here is that the magnetic orientations of the 
individual monopoles appear as 
 independent, unconstraint 
coordinates in the moduli space.
This appears to support  Murray's interpretation  \Murray\ that,  
at least in a suitable asymptotic region, 
 $M_{2,0}$ describes two monopoles with six independent dynamical
degrees of freedom each. Leaving aside for a moment the difficulties
of talking about the individual monopole charges in a multimonopole
configuration (we return to this point in Sect.~9) it is clear that 
two charge one  monopoles with arbitrarily oriented 
vector magnetic charges would combine to a monopole configuration
 whose vector
magnetic charge in general violates the Dirac condition. 
In the next two sections we will give a number of arguments why
the interpretation of the space  $M_{2,0}$ in terms of two
monopoles with independent dynamical vector magnetic charges is
not correct.  Instead the dimensionality of $M_{2,0}$
can be understood by taking into account a   new dynamical 
coordinate which only appears when two monopoles are combined into
a monopole configuration of topological charge two.  
 
\newsec{Dancer's moduli space }

A detailed study of the space $M_{2,0}$, including its Riemannian
structure,  was carried out by Dancer
in a series of papers \And\ -\Anddd\ from the point of view of 
Nahm matrices; see also the papers \DL\ and \DLL\
 with Leese, where the classical
dynamics of charge two monopoles is studied.
For us the description of the isometries of $M_{2,0}$ in  \Andd\ 
is particularly relevant. There it is shown that $M_{2,0}$ is 
a hyperk\"ahler manifold whose double cover 
decomposes as a direct
product of hyperk\"ahler manifolds  such that one  has the isometry
\eqn\mtwelve{
 M_{2,0} = \bR^3 \times{S^1 \times \tilde M^8 \over \bZ_2},
}
where $\tilde M^8$ is  an eight-dimensional  
irreducible hyperk\"ahler manifold.
The translation group acts only  on the $\bR^3$ part  of the above 
decomposition and the central $U(1)$ subgroup  of the unbroken gauge group 
$U(2)$ acts only on the $S^1$ factor. 
Thus we may think of $\tilde M^8$ as the space of centred 
$SU(3)$ monopoles (in analogy with the space
$M_2^0$ in \sutwo\ for $SU(2)$ monopoles). 
The manifold  $S^1 \times \tilde M^8$ 
is acted on isometrically by Spin$(3)\times
SU(2)$, where Spin$(3)$ is the double cover of the group $SO(3)$
of spatial rotations and $SU(2)$ is the quotient group of
the unbroken gauge group $U(2)$ by its centre $U(1)$.  
The centre of $SU(2)$ acts trivially but the centre of Spin$(3)$ acts
non-trivially on both  $\tilde M^8$ and on  $S^1$, the action
on the latter being rotation by $\pi$.

On the quotient $M^8 =\tilde
 M^8/\bZ_2$ the Spin$(3)$  action descends to an 
 $SO(3)$  action which commutes with  the  $SU(2)$ action. 
Thus by quotienting $M^8$ further
by the $SU(2)$ action one obtains
a five-dimensional manifold $N^5$ which is acted on isometrically
by $SO(3)$.   Physically this space describes monopoles with
fixed centre-of-mass, quotiented 
by the action  of the unbroken gauge group. 
It turns out to have  simple geometrical interpretation.
In \Andd\ it  is explained that $N^5$ can be identified with a certain open
subset of the space of symmetric traceless $3\times 3$ matrices,
with $SO(3)$ acting by conjugation. We shall now show that 
one can further associate a unique unoriented ellipse or line segment 
in $\bR^3$  to a given traceless symmetric matrix. This point
of view is particularly convenient for us because it
exhibits very clearly the orbit structure of the $SO(3)$ action.

Given a traceless symmetric matrix,
 diagonalise it and order
the eigenvalues $\lambda_+ \geq \lambda_0 \geq \lambda_-$.
In the generic case $\lambda_+ > \lambda_0 > \lambda_-$
call the associated eigenvectors $v_+, v_0, v_-$ respectively
and define an unoriented  ellipse in the $v_+,v_0$ plane whose major
axis is along $v_+$ and has length  $A = \lambda_+ - \lambda_-$
and whose minor axis is along $v_0$ with length 
 $B = \lambda_0 -\lambda_-$.
When $\lambda_+> \lambda_0 =\lambda_-$ this degenerates to a line along 
$v_+$ with length $A = \lambda_+ - \lambda_-$  and when
$\lambda_+ =  \lambda_0  > \lambda_-$ it becomes a circle orthogonal
to $v_-$ with radius  $A=B= \lambda_0 -\lambda_-$. In either case 
the coincidence of  two eigenvalues means that the matrix
is invariant under conjugation by some  $O(2)$ subgroup 
of $SO(3)$   and this  invariance 
is reflected in the  axial symmetry of the associated 
figure. Finally in the 
completely degenerate case $\lambda_+ = \lambda_0 =\lambda_- = 0$
the associated ellipse degenerates to a point, so both
the matrix and the associated figure  are   kept fixed
by  the $SO(3)$ action.

When one takes the quotient of the space $N^5$ by the $SO(3)$ action
one obtains a set $N^2$ which is
 not a manifold because the isotropy group  is not the same 
at all points of $N^5$. Nonetheless the set $N^2$ is very interesting 
to consider: it contains those parameters in $M_{2,0}$ which cannot
accounted for by actions of the symmetry group and which
therefore may be thought of as irreducible ``shape'' parameters
of the monopoles.
 In \Andd\ it is shown  that
\eqn\ntwo{
N^2 = \{(D,\kappa): 0\leq \kappa \leq 1, 0\leq D < {2\over 3} E(\kappa)\}, 
}
where $E(\kappa)$ is the elliptic integral
\eqn\ellipint{
E(\kappa) = \int_0^{\pi\over 2} { d \theta \over 
\sqrt{1- \kappa ^2 \sin^2 \theta}}.
}

In our description of $N^5$ in terms of  ellipses
we also isolated  shape parameters, namely the lengths $A$ and $B$ of the 
major and minor axes. 
Using the explicit  map between   Nahm data and traceless
symmetric matrices given in \Andd\ we can write down the relation
between these and the coordinates  $(D,\kappa)$ for the space $N^2$:
\eqn\ellin{
A= {1\over 2} D^2 \quad \hbox{and} \quad B= {1\over 2} (1-\kappa^2) D^2.
}
In Fig.~2 we sketch the parametrisation of the space $N^2$
in terms of $A$ and $B$.
Using further  the map between Nahm data and monopoles we can now 
in principle establish a correspondence between  ellipses in $\bR^3$
and  monopoles.
In practice detailed information about the monopole fields
is only available  for particularly symmetric configurations.
It is these which we will briefly review. 

\vskip .5in
\input epsf 
\epsfxsize=16truecm
\centerline{
\epsfbox{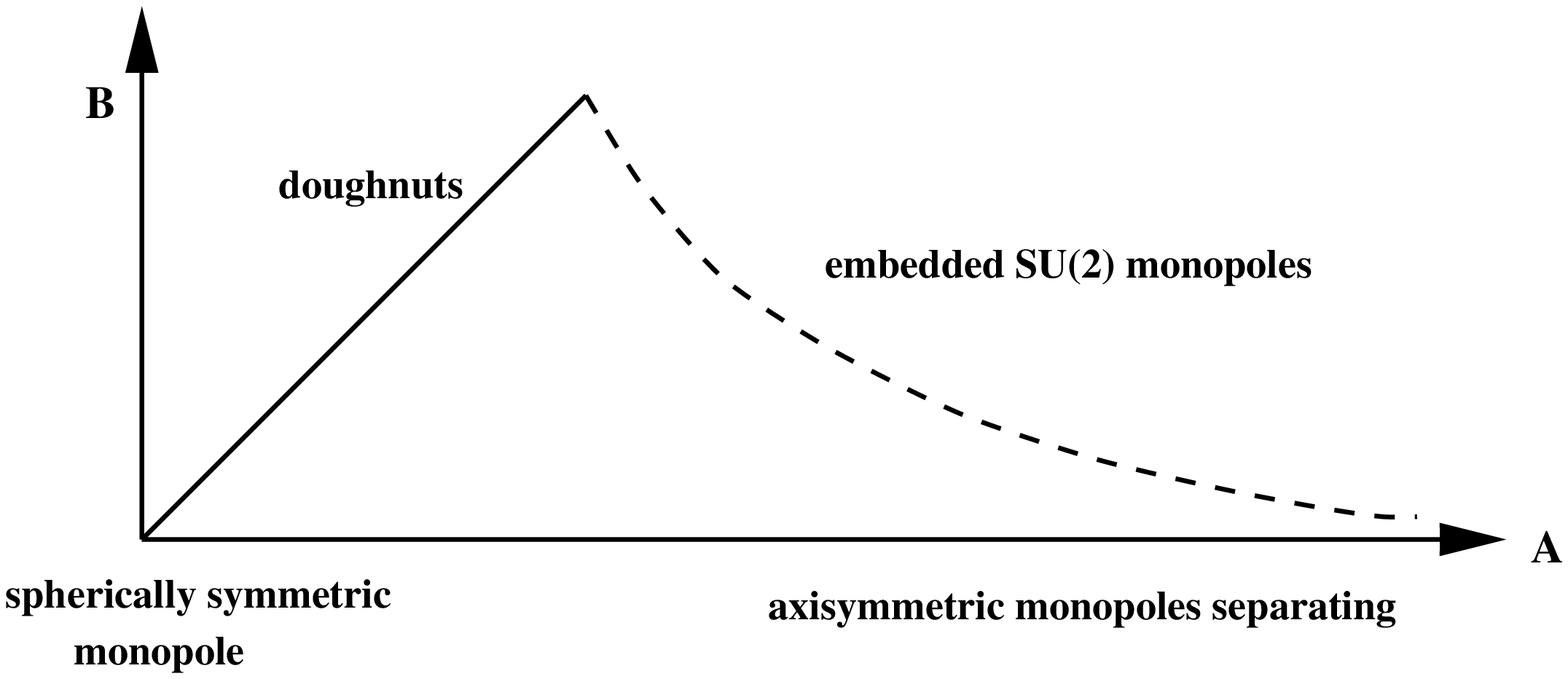}
}
\bigskip
{\footskip14pt plus 1pt minus 1pt \footnotefont
\centerline{\bf Fig.~2 }
\centerline { \it Monopole configurations parametrised by $N^2$}
}
\bigskip

 The point $A=B=0$  where the corresponding ellipse 
becomes a point 
represents  the spherically symmetric monopole
(in the sense that a spatial $SO(3)$  rotation acting on such 
a monopole is equivalent to an internal $SU(2)$ transformation).
The precise functional form of such spherically symmetric monopoles
was first studied in \BW.
The line defined by  $\kappa=0$ 
corresponds, in the ellipse picture, 
 to a  family   of circles with radius $0\leq A=B < 2E(0)/3 =\pi/ 3$
 and represents axisymmetric monopoles,
called trigonometric
axisymmetric in \Andd. They  are in fact the family of solutions
found by Ward in \Rich\ and   have the energy concentrated
in a doughnut-shaped region, with the maximum of the energy density
on a circle. This family approaches the embedded unique axisymmetric
charge two $SU(2)$ monopole  as $D\rightarrow \pi/3$. As we  know from 
previous sections that  embedded solution is an element of the small
stratum $M_{2,1}$. More generally,  for each  $0\leq \kappa < 1$, 
configurations in $N^2$ approach embedded $SU(2)$ monopoles belonging
to $M_{2,1}$ in  the limit
$D \rightarrow E(\kappa)$. 
The condition $\kappa=1$  also defines a line in $N^2$
which corresponds 
to degenerated ellipses of vanishing minor axis $B$ and arbitrary 
length of the major axis $A$
(for $\kappa \rightarrow 1 $ the integral defining $E(\kappa)$ diverges
logarithmically).
This line also represents axisymmetric monopoles,
called hyperbolic axisymmetric in \Andd. The functional
form and energy distribution of these monopoles along the axis 
of symmetry is given in  \And. 
Essentially, the family of hyperbolic axisymmetric
monopoles interpolates between the spherically symmetric
charge two monopole and  configurations consisting of two well-separated
charge one monopoles, with the separation approximately given
by $D$.

Unfortunately, little is known about  the fields of
axisymmetric monopoles off the axis of symmetry and about 
the  monopoles represented
by a generic point in $N^2$ (see, however, \DL\ and  \DLL\ for numerical
information).
The study of the axisymmetric solutions suggests 
 that one of the parameters in $N^2$ should 
be thought of as a separation parameter, and that, in the ellipse 
parametrisation, that separation is given in terms of 
the major axis as $D=\sqrt {2 A}$. Then there is a complementary
parameter --- in the ellipse picture we think of the length of 
the minor axis --- which parametrises some kind of internal 
deformation of the monopoles. This parameter corresponds to 
what in the recent literature \LWY\  has been called  the 
`non-abelian cloud' of the monopole. It is the goal of 
the next section to make that notion more precise.

\newsec{Monopoles and non-abelian dipoles}

It is known \MS\ that  charge two monopole solutions in 
spontaneously broken $SU(2)$  Yang-Mills-Higgs theory have 
no magnetic dipole moments. This is easy to understand when the
two monopoles are separated:  the  charges are equal, and reflection
symmetry forces dipole moments relative to the centre-of-mass to be 
zero. The monopoles in minimally broken $SU(3)$ gauge theory, however,
carry vector magnetic charges and two monopoles carrying
the same topological magnetic charge may carry different non-abelian 
magnetic charges. It is then natural to expect 
multi-monopole configurations
made up of two or more such monopoles to have non-vanishing
non-abelian magnetic dipole moments. What is more, in view of 
the difficulties of assigning, even in principle,
 individual vector magnetic charges
to monopoles in a multi-monopole configuration,  dipole moments
(and possibly higher multipole moments) appear to be the only
source of information about the magnetic charges of the individual
monopoles. This is the point of view we will adopt in this section.

All the  fields we 
study here have Higgs fields whose asymptotic expansion is consistent
with  the general form
\eqn\dipex{
\Phi(\br) = \Phi_0(\hat{\br}) - { G_0(\hat{\br})  \over 4 \pi r}
+ { i d^{aj}\,I_a(\hat{\br})\, \hat{r}_j \over e^2 r^2} + {\cal O}
\left({1\over r^3}\right)
}
with summation on repeated indices.
Here $\Phi_0(\hat{\br})$ and $G_0(\hat{\br})$ are the vacuum expectation
value of the Higgs field and the vector magnetic charge on the 
two-sphere at infinity in some gauge (for configurations
in $M_{2,0}$,  $\Phi_0(\hat{\br})$ and $G_0(\hat{\br})$  are parallel
everywhere), and   $I_a(\hat \br)$, $a=1,2,3$, 
 are the generators of the unbroken
$SU(2)$ gauge group at a point $\hat \br$
on the two-sphere at infinity (since $K$ is even there is no topological
obstruction to writing them down on the entire two-sphere
at infinity). 
Without entering a general discussion of  multipole expansions
of non-abelian gauge fields we interpret the $(1/r^2)$-terms  in the above
expansion as a dipole term. The Bogomol'nyi equation relates 
the $(1/r^2)$-terms in the Higgs fields with $(1/r^3)$-terms in the 
magnetic field and we may thus consider the coefficient matrix
$d^{aj}$, $a,j=1,2,3$, which 
transforms as a vector both under spatial rotations and under
rigid $SU(2)$ gauge transformations, as a non-abelian
magnetic dipole moment.

We are interested in the dipole moments of 
 monopole configurations  in $M_{2,0}$, but
unfortunately, explicit computations of 
the Higgs  field for a generic   configuration in this space have
only been carried by a numerical implementation of the Nahm transformation
  (see  \DL\ and \DLL).  
The direct extraction of the dipole moments from the (explicitly
known) Nahm data seems very difficult so we  restrict our 
discussion to configurations with extra symmetries or to  certain
limits, where analytic expressions for the Higgs field
 have been given in the literature. The configurations we shall 
discuss are essentially those which are on  or near the boundary
of the set $N^2$ depicted in Fig.~2. 

We begin with  hyperbolic  axisymmetric monopoles. The Higgs field
on the axis of symmetry, taken to be the $z$-axis, was written down
in \And, and extracting the $(1/z^2)$-term we indeed find a non-abelian 
component: 
\eqn\hypasy{
\Phi(0,0,z) = \left(1-{1\over 2 e  z}\right) \Phi_0
+ {i D \coth 3 D\over 2  e^2  z^2} \,\, I_3 + {\cal O}
\left({1\over r^3}\right). 
}
Since we do not know the Higgs field along any ray other than the 
$z$-axis we  cannot deduce the full tensorial structure of the 
dipole moment. However, we conjecture
that for a hyperbolic  axisymmetric monopole configuration
consisting of two monopoles well-separated    along the 
$z$-axis   the dipole moment is of the form $d^{aj}= 
d \delta^{a3}\delta^{j3}$.
From \hypasy\ the  coefficient $d$  is  then $D \coth 3 D /  2 $.
For large $D$, $D \coth D \approx D$, so the dipole moment
 increases linearly with separation.
This is precisely the dipole moment  one  expects for a configuration
containing two monopoles, separated along the $z$-axis,
 with equal and opposite non-abelian magnetic charges.

In the limit $D\rightarrow 0$  the  monopoles coalesce to the 
spherically symmetric solution. Remarkably the dipole moment 
does not vanish in this limit.
Exploiting the spherical symmetry we can now write down the 
asymptotic form of the solution everywhere. In the so-called
singular gauge, where the gauge-potential has a Dirac-string
singularity, it has the form
\eqn\sphereasy{
\Phi(\br) = \left(1-{1\over 2 e  r}\right) \Phi_0
+ {i\over 6 e^2 r^2}\,\, \bI \cd \hat \br  + {\cal O}
\left({1\over r^3}\right) . 
}
The non-abelian  dipole moment is thus seen to be of the 
hedgehog form $d^{ai} =d\delta^{ai}$:
 the Lie algebra components
of the dipole are correlated with their spatial direction in such
a way that the non-abelian dipole moment is invariant under 
simultaneous spatial rotations and global $SU(2)$ gauge rotations.
In particular the dipole moment therefore does not single out
a preferred spatial direction.

Moving on to  trigonometric axisymmetric monopoles
we find  the following
expansion  along the axis of symmetry, taken to be the $z$-axis:
\eqn\trigasy{
\Phi(0,0,z) = i\left(1-{1\over2 e  z}\right)\Phi_0
+ {i D \cot 3 D\over 2   e^2 z^2}\,\, I_3  + {\cal O}
\left({1\over r^3}\right).
}
Again we are unable to deduce the  tensorial structure
of the non-abelian dipole moment, but 
we can determine  the component  $d^{33} =  \cot 3 D / 2$.
  Since all trigonometric axisymmetric monopoles have their energy density
concentrated in a finite region of space, their dipole moments
cannot be understood in terms of the separation and individual
non-abelian monopole charges of  two single monopoles. Rather,
the above expansion shows that 
trigonometric axisymmetric monopoles have an intrinsic non-abelian 
dipole moment. In the limit  $D\rightarrow \pi/3$, where
the  monopole configurations  approach 
 the  toroidal  charge two monopole 
in the small stratum, 
the dipole strength goes to infinity, giving a very physical
interpretation of the transition between the strata: the
infinite non-abelian  dipole 
moment   combines with the (essentially abelian)  vector  
magnetic charge of the large stratum
diag$({1\over 2},{1\over 2},-1)$
to produce the vector  magnetic charge diag$(0,1,-1)$ characteristic 
of the small stratum. 

Finally we turn to the asymptotic form of configurations 
near the boundary of $N^2$ drawn as a dashed line in Fig.~2.
 This is  studied 
in  \irwin\ where  it is argued that for these  monopole configurations
the   term in the Higgs field  which we call the  dipole term 
is again of the hedgehog  form:
\eqn\cloudasy{
\Phi(\br) = i\left(1-{1\over 2 e z}\right) \Phi_0
- {i D \over 2(\pi -3 D) e^2 z^2}\,\,\bI \cd \hat \br  + {\cal O}
\left({1\over r^3}\right).
}
Note that near the boundary the coefficient$ D /2(\pi -3 D)$
is necessarily large and tends to infinity as $D\rightarrow \pi/3$.

Based on  the sample of configurations studied here we
 propose that  quite generally  the parameters in $N^2$
can be understood physically  as  characterising 
the monopoles'  non-abelian dipole moments.
Configurations on the line $\kappa=1$  which consist of two well-separated
monopoles
have a  purely extrinsic dipole moment. This results from the individual
monopoles' opposite  non-abelian magnetic charges and is  essentially
a measure of the monopole separation. Configurations away from the line 
$\kappa =1$ have dipole moments which cannot be understood in
terms of the   individual  magnetic charges. In particular
the fact that  configurations made up of two well-separated
monopoles have dipole moments of the hedgehog type \cloudasy\
in a certain limit shows that the  fields of the two monopoles which make 
up such configurations must be quite different from the fields of 
single monopoles in $M_{1,1/2}$. The asymptotic form \cloudasy\
rules out the possibility, suggested by the rational map description,
that in $M_{2,0}$ two  well-separated monopoles have
magnetic charges with independent directions. Instead it suggests
that such monopoles still have their magnetic charges anti-aligned,
but that in addition they  have individual dipole moments of the hedgehog type.
Thus we propose that  the elusive cloud parameter is 
a measure of the individual
monopoles' dipole strength. In the limit $D \rightarrow E(\kappa)$
this dipole strength  tends to infinity. The total
dipole moments is then also of the hedgehog type since, for fixed 
separation,  the 
extrinsic dipole moment   is  negligibly small  relative to
 the individual hedgehog dipole moments in this limit.

Armed with a physical interpretation  of the moduli in $M_{2,0}$
we now turn to the quantisation of monopoles. There,  the conclusion that
in any configuration in $M_{2,0}$
which consists of well-separated monopoles the 
individual monopoles' non-abelian charges  are  anti-parallel
will be crucial.

\newsec{Monopole quantisation, the dyon  spectrum and the emergence  of
 $U(2)\semidir \bR^4$}

The quantisation of 
BPS monopoles in Yang-Mills-Higgs theory with gauge group $SU(2)$
 has been studied extensively in the literature. In that case 
the moduli spaces are smooth manifolds with finite 
metrics (and hence well-defined integration measure)
and in the standard bosonic quantisation scheme 
one takes the Hilbert space of states to be the set of all square-integrable
functions on the  moduli space, see e.g. \GM\  and \berni.
This quantisation scheme needs to be amended before it can
be applied to  the $SU(3)$ monopoles studied here: the smallest  strata of 
the moduli spaces, for example, are fibred over a two-sphere \fibre\
(the magnetic orbit)  which inherits from the field theory 
 neither a metric nor a measure nor even a differentiable structure.
The largest strata (for even $K$),
 on the other hand, are smooth manifolds  with finite
metrics and the above scheme can again be applied without difficulty.
Our prescription will have to take these differences into account.

Our basic starting point is that magnetic charges label superselection
sectors in  quantised Yang-Mills-Higgs theory. 
The rationale here is that magnetic charges are conserved independently
of the equations of motion and that  there are no finite-action 
configurations which interpolate between classical configurations 
with different magnetic charges. Thus there can also be no
quantum mechanical transitions between states with different
magnetic charges.
This superselection rule is implicit in the standard 
quantisation scheme applied to monopoles in theories with
abelian unbroken gauge group. There all the magnetic charges
are topological and label disjoint moduli spaces. 
Superpositions between states of different magnetic charges
would correspond to linear combinations   of  wavefunctions 
on different moduli spaces, and these are usually ruled out.
In our case, the 
superselection sectors are labelled by  pairs $(K,\bk)$ of
topological and non-abelian magnetic charges.
If $K$ is even and 
the non-abelian magnetic charge vanishes the corresponding
sector takes the familiar form:
it is given  by the Hilbert space of square-integrable
functions on the moduli space $M_{K,0}$ which, being  the  largest
stratum for given $K$,  is equipped with a smooth and finite metric.
In symbols
\eqn\hilb{
{\cal H}_{K,0} = L^2( M_{K,0}).
}
If $k>0$ on the other hand, we label sectors also by a point $\bk$ on
 the magnetic orbit and define it to 
be the Hilbert space of square-integrable functions on the corresponding
 fibre $M_{K,\bk}$ defined  after \fibre\ for the case $k=K/2$
(as also noted there, however, we expect no 
principal difficulty in defining the fibres $M_{K,\bk}$ for any 
of the allowed pairs $(K,k)$).  Since these fibres
are again equipped with  finite and smooth metrics this definition
makes sense. However, there is another way to define the superselection
sectors labelled by  $(K,\bk)$ which is more useful for us.
The starting point here is  interpretation  in 
the second but last paragraph of Sect.~5  of the 
base spaces of the fibration $M_{K,k} \rightarrow S^2$ as  round  spheres
with integration measure
 $k^2 \sin \beta d\beta\wedge d \alpha$. This allows us 
to define a measure on $M_{K,k}$, namely the product measure of 
$k^2 \sin \beta d\beta\wedge d \alpha$ with the natural measure on
the fibres (the one coming from the metric induced by  the field theory).
Then it makes sense to define the Hilbert space
\eqn\hilbbb{
{\cal H}_{K,k} = L^2(M_{K,k}). 
}
From this larger space we can project out the  desired superselection
sectors as follows. Consider the operator
\eqn\clevv{
p_i:{\cal H}_{K,k} \rightarrow {\cal H}_{K,k} 
}
whose action on a  function $\phi\in {\cal H}_{K,k}$
is 
\eqn\clebv{
 p_i\circ \phi (X) = k_i \phi(X),
}
with $\bk =\pi^k(X)$ in terms of the projection map $\pi^k$
defined in \fibre\ for $k=K/2$ but conjectured to exist for 
all $k>0$ (in the allowed range) in Sect.~5.
The simultaneous eigenspaces ${\cal H}_{K,\bk}$
of the three operators $(p_1,p_2,p_3)$
with eigenvalues $(k_1,k_2,k_3)$
contain precisely  the states with definite magnetic charge $\bk=(k_1,k_2,k_3)$
and thus correspond   to the required superselection sectors.
As we shall see, this construction has a very natural interpretation later
in this paper.

Having defined the Hilbert spaces for our quantisation scheme
we turn to the actions of the various symmetry groups on these 
spaces.
Generally speaking the symmetry group of the theory acts on the 
moduli spaces, and hence also on functions on the moduli spaces, 
which can therefore be organised into representations of the 
symmetry group. In our case the spatial symmetry groups
of  translations and rotations act
smoothly  on the moduli spaces, and wavefunctions can correspondingly
be organised into momentum and angular momentum eigenstates. 
 Here we are only interested in the transformation
properties of  of wavefunctions
under the action of the unbroken gauge group $U(2)$. As explained 
in Sect.~5
this action is smooth and isometric only on the largest stratum
of moduli spaces for monopoles with even topological charge. On all
other strata it is obstructed by the non-abelian magnetic charge
so that only the centraliser of the  vector magnetic charge
acts smoothly and isometrically. Our superselection
sectors reflect that difference.
 Thus we expect to be able to organise the elements of
${\cal H}_{K,0}$ into representations  
of $U(2)$ while elements of ${\cal H}_{K,\bk}$  
should be organised into  representation 
of the centraliser of $\bk$. In the former case we may therefore 
denote dyonic quantum 
states by $|K,0; N,j,m \rangle$, where $K$ is the 
topological magnetic charge, $N$ is an integer which specifies the 
representation of the $U(1)$ transformation generated by  $Y$
\ugen\ and  the half-integers $j\in \{0, {1\over 2}, 1, {3\over 2} ...\}$, 
 $m\in \{ -j,-j+1, ..., j-1,j\} $  specify a state
 $|jm\rangle$ in an  $SU(2)$ representation; the three numbers $(Njm)$
then specify a state in an  $U(2)$ representation, 
and we need to impose the 
$\bZ_2$  condition that $N$ is odd if $j$ is a  half-odd integer
and $N$ is even if $j$ is an integer. The other sort of dyonic states
can be written as $|K,\bk; N,s \rangle$, where  $K$ and $N$ have 
the same meaning as before,  
 $\bk$ is the non-abelian magnetic
charge defined in \magvec,  and  $s$
is a single half-integer which  specifies
the representation of the $U(1)$ subgroup of $SU(2)$ which leaves
$\bk$ fixed. Again we have the  condition that $N$ is odd
if $s$ is half-odd and $N$ is even if $s$ is integer. 

We now show how these states are realised as wavefunction
on the moduli spaces. 
Since all the essential features of the scheme we are going to
propose show up already for monopoles of topological charge $K\leq 2$,
 we will restrict attention to the corresponding
moduli spaces in the following. Thus we will be able to draw on
the explicit description of those spaces in the previous sections.

Again we  begin with a single monopole and the moduli 
space $M_{1,1/2} = \bR^3 \times S^3_P $. As explained in sects. 5 and 6,
this space is fibred over the magnetic orbit $S^2$.
A point $\bk$ on the magnetic orbit specifies the non-abelian
magnetic charge and obstructs the $U(2)$ action: only the 
centraliser acts smoothly.
For an explicit description of quantum states of a single monopole
it is convenient to write a generic $SU(2)$ matrix  
in terms of 
Pauli matrices and Euler angles $(\alpha,\beta,\gamma)$ 
as in \qmat\ and to 
introduce Wigner functions $D_{ms}^j(Q)  = 
e^{-im\alpha} d^j_{ms}(\beta) e^{-is \gamma}$
on $SU(2)$ following the conventions of \threerus.
The functions $D_{ms}^j$, $ j \in \{ 0, {1\over 2},1, {3\over 2}, ...\}$,
$m,s \in \{ -j,-j+1, ... , j-1, j \}$ form a basis for the Hilbert space
$L^2(SU(2))$, and one can moreover (formally) expand the $\delta$-function
on $SU(2)$  in terms of them. We write the $\delta$-function peaked at the 
 point $  Q$ and with argument $Q'$ as   $\delta_{E}(Q'Q^{-1})$,
where $E $ is the identity element. Then, putting primes on the Euler angles
for $Q'$ we have the formula 
\eqn\delfun{
\delta_{E}(Q' Q^{-1}) = \delta(\alpha'- \alpha)
\delta(\cos \beta' - \cos \beta) \delta (\gamma' - \gamma)
}
and the expansion
\eqn\delfunex{
\delta_{E}(Q' Q^{-1}) =
\sum_{j=0,{1\over 2}, 1 ...} \sum_{m=-j}^j \sum_{s=-j}^j
{2j+1 \over 16 \pi^2} D^{*j}_{ms}(Q') D^j_{ms}(Q ).
}

Quantum states for a single monopole are of the form 
$|1, \bk; N, s\rangle$, where $|\bk|=1/2$ and we have the 
constraint $N+2s =0$ as a consequence of the monopole's invariance under 
the generator $Y+2I_3$.  Since the 
point $\bk$  on the magnetic orbit is  obtained from a general point $Q$
on $SU(2)$ via the Hopf projection we have the explicit parametrisation
 in terms of Euler angles $\hat \bk =(\sin \beta \cos  \alpha,
\sin \beta \sin  \alpha,
\cos   \beta)$. Furthermore we can give an explicit
realisation  of  the states  of a single monopole  
in terms of Euler angles:
\eqn\hels{
\langle \chi',\alpha', \beta', \gamma' |1, \bk;N, s\rangle = 
e^{i N \chi'}
\delta(\alpha' -\alpha) \delta(\cos \beta' - \cos \beta) e^{is \gamma'},
}
where the condition $N=-2s$ ensures that the right hand side 
only depends on $2\chi' - \gamma'$. 
Then, using  the above completeness relation \delfunex\ we also deduce
\eqn\helss{
 \langle \chi', \alpha', \beta', \gamma' |1,\bk;N, s\rangle =
e^{iN \chi'} 
\sum_{j \geq |s|} \sum_{m=-j}^{m=j}{2j+1\over 4 \pi}
D^{j*}_{ms}(Q') e^{-im \alpha} d^j_{ms}(\beta).
}

Turning next to two monopoles, we have to distinguish the 
two strata. Quantum states on the small  stratum are again much
like the quantum states of a single monopole. They are  of the 
form $|2, \bk; N, s\rangle$, where now $|\bk|=1$ and we again have the 
constraint $N+2s =0$. These states can be realised as  wavefunctions
on the moduli space $M_{2,1}$. The most general wavefunction
will also depend
on the centre-of-mass position and on the relative coordinates 
summarised in the Atiyah-Hitchin manifold $M_2^0$, but here we are only
interested in the  dependence on $S^3_P$. This dependence is just 
like for the quantum states of a single monopole:
\eqn\helsss{
 \langle \chi', \alpha', \beta', \gamma' |2,\bk;N, s\rangle =
 e^{iN \chi'}  
\sum_{j \geq |s|} \sum_{m=-j}^{m=j}{2j+1\over 4 \pi}
D^{j*}_{ms}(Q') e^{-im \alpha} d^j_{ms}( \beta), 
}
where now $|\bk|=1$.
There may appear to be an inconsistency in our notation at this 
stage, in that the magnetic labels appear on the right  hand
side of \helss\ and \helsss, but not on the left. 
However, the rationale for our notation will soon become clear.

The large stratum $M_{2, 0}$ is smooth with a finite metric
and, like in the case of $SU(2)$ monopoles, quantum states can
be realised as smooth wavefunctions on the moduli space. The  magnetic
orbit is trivial in this case, so all coordinates should be treated 
on the same footing. The most general 
wavefunction  depends on the $U(2)$ Euler angles
$(\chi,\alpha,\beta,\gamma)$, on the centre-of-mass
position, on  spatial Euler angles
 and on the shape parameters $(A,B)$.
A detailed investigation of the quantum mechanics on $M_{2,0}$
would be a very interesting but also very challenging project.
Here we are merely interested in the transformation properties of
the wavefunction under the unbroken gauge group $U(2)$ so we 
focus again on the wavefunction's dependence on the corresponding
 coordinates. Making use of the fact that, for any fixed $s$ 
in the allowed range, both the Wigner functions 
$D_{ms}^{j}$ and their complex-conjugates $D_{ms}^{j*}$, 
$m \in \{-j, -j+1, ..., j-1, j\}$
 span the spin $j$ representation of $SU(2)$, we may choose a value for 
$s$ and then represent dyonic quantum states as follows:
\eqn\utwost{
\langle \chi' , \alpha', \beta', \gamma' |2,0;N, j,m \rangle_{s}=
{\sqrt{2j+1} \over 4 \pi}
e^{iN\chi'} D_{ms}^{j*}(Q').
}
Different values for $s$  lead to equally valid realisations
of the state $|2,0;N, j,m \rangle$, but we introduce the 
suffix $s$ here because we shall see further below 
 that this half-integer also has a physical interpretation.

Having represented dyonic  quantum states with  
topological  magnetic charges
one and two  as wavefunctions on the relevant moduli
spaces we are in a position to address one of the key concerns 
of this paper: what is the relationship between the tensor
product of two  quantum states
of a single monopole and  a quantum state of a 
topological charge two monopole? The answer to
this question can in principle
be deduced from a knowledge  of the algebraic object whose representations
are being studied. In our case we do not yet know the 
relevant  algebraic object.
It is clearly not simply the unbroken gauge group $U(2)$ since
the quantum states we have studied carry in general 
 only a representation of a 
subgroup of it, namely the centraliser of the 
non-abelian magnetic charge. 
The algebraic object we seek should incorporate the subtle
interplay between magnetic and electric properties which
we have seen in the dyonic states discussed so far.
We will now argue that for 
dyonic states in the present theory 
the relevant algebraic object is the semi-direct product 
$U(2) \semidir \bR^4$. Denoting the generators of $U(2)$ again
by  $I_1,I_2,I_3$ and $Y$ and calling the translation generators
 $p_1,p_2,p_3$ and $P$ the Lie algebra of $U(2) \semidir \bR^4$ has 
 the following commutation relations
\eqn\comrel{\eqalign{
[I_a,I_b] &=i\epsilon_{abc}I_c \cr
[p_a,p_b] &=0  \cr
[I_a,p_b] &=i\epsilon_{abc}p_c \cr
[Y,P]=[Y,I_a]=[Y,p_a]=[P,I_a]&=[P,p_a] = 0, \,\,\hbox{for}\,\, a=1,2,3.
}}
 To test this proposal we first show
that the representations contain precisely the  
dyonic states discussed above and then show that the Clebsch-Gordan
coefficients of $U(2) \semidir \bR^4$ 
are consistent with the representation of states as wavefunctions
on moduli spaces.

The semi-direct product   $U(2) \semidir \bR^4$ is an example of 
a regular semi-direct product and therefore its representation 
theory is most effectively dealt with via the method of induced 
representations \indrep. For a general regular semi-direct product 
$S= H \semidir N$, where $N$ is an abelian, normal subgroup of $S$,
the construction of an irreducible representation begins with the 
classification of all $H$ orbits $O$  in $\hat N$, the set of characters
of $N$. Each orbit has a characteristic centraliser group $C$ (the 
centraliser group of any point on it) and  a unitary irreducible 
representations
(UIR) can then be induced from the group $C \semidir N$. The representation
space $V_{O,\rho}$ of an  UIR  of $S$ constructed
in this way is then labelled by the orbit $O$ and a UIR $\rho$
of the stability group $C$:
\eqn\genin{\eqalign{
V_{O,\rho}=\bigl\{\phi:H \rightarrow V_{\rho}\big \vert
\phi(QX) =& \rho(X^{-1})\phi(Q)\,\, \forall Q\in H,X\in C \cr
 &\hbox{and}\,
\int_{H/C} ||\phi||^2(z) d\mu(z) < \infty \bigr\},
}}
where $V_{\rho}$ is the carrier space of the representation $\rho$,
$||\cd||$ is the norm induced by the inner product in $V_{\rho}$
and $d\mu$ is an invariant measure on the coset $H/C \cong O$. 
One can show that all UIR's of $S$
can be obtained in this way. 

Applying this theory to our case we 
first note an obvious simplification.  Since $U(2) =
\bigl( U(1) \times SU(2) \bigr) /\bZ_2$  we have a corresponding 
direct product decomposition $U(2) \semidir \bR^4 = 
\bigl( (U(1)\semidir \bR) \times (SU(2)\semidir \bR^3) \bigr)/\bZ_2$. 
The interesting part
of the representation theory comes from the non-abelian part
 $ SU(2)\semidir \bR^3$, which happens to be the double cover of the 
Euclidean group in three dimensions and whose representation theory is
particularly well documented in the literature. In this case the set
of characters  $\hat \bR^3$ of the translation group $\bR^3$ is isomorphic to 
$\bR^3$ and can  physically  be thought of as momentum space, with elements
denoted by $\bk$. The $SU(2)$
orbits in $\hat \bR^3$ are spheres with radius $k>0$ or simply
a point. In the former case the centraliser  group is $U(1)$ and in the latter
the entire group $SU(2)$. By the general theory we thus obtain
two sorts of UIR's  of $ SU(2)\semidir \bR^3$. If $k > 0$
the carrier spaces
are infinite dimensional and 
  labelled by the orbit size $k>0$ and a half-integer $s$
specifying a $U(1)$ representation. They may be realised as follows
\eqn\carry{\eqalign{
V_{k,s} = \bigl\{ \phi: SU(2) \rightarrow \bC \,\big \vert
 \phi(Qe^{-{i\over 2}\xi\tau_3})
=& e^{i s\xi} \phi(Q)\cr
& \hbox{and}\, \int_{S^2} 
|\phi|^2(\alpha,\beta) \sin\beta\, d\beta d\alpha < \infty\bigr\}.
}}
The action of an element $(A,\ba) \in SU(2) \semidir \bR^3$
on this Hilbert space is given by
\eqn\acton{
((A,\ba)\circ \phi)(Q)= e^{i\ba\cd\bk}\phi(A^{-1}Q),
} 
where $\bk$ is obtained from $Q$ via the Hopf projection \hopfi\ i.e.
$\bk = \pi_{\footnotefont{Hopf}}^k(Q)$.
Having earlier introduced   Wigner functions $SU(2)$  as  
 a basis of $L^2(SU(2))$  one checks   that for fixed $s$
the functions  $D^{j*}_{ms}$, $
j\in \{|s|, |s|+1 , ...\}$,$ m\in\{-j,-j+1, ..., j-1,j\}$ form a basis of
$V_{k,s}$. 

In the case $k=0$ the general prescription of induced representation
theory starts from a standard $(2j+1)$-dimensional representation 
of $SU(2)$ with carrier space  $\bC^{2j+1}$  on which $Q\in SU(2)$
is represented by the matrix $D_{ms}^j(Q)$, $s,m \in
\{ -j, -j+1, ..., j-1,j\}$,
and  leads to the Hilbert space
\eqn\carmo{
V_{0,j} = \bigl \{\phi: SU(2) \rightarrow \bC^{2j+1}\big\vert \phi_s(QQ')=
\sum_{t=-j}^j D_{st}^j(Q'^{-1}) \phi_t(Q) \bigl\},
}
where $\phi_s$ denotes the $s$-th component of $\phi$.
The action of  $(A,\ba) \in SU(2) \semidir \bR^3$
reduces to an $SU(2)$ action:
\eqn\acton{
((A,\ba)\circ \phi)(Q)=\phi(A^{-1}Q),
} 
Again it is straightforward to write down a basis. 
Since the right action of $SU(2)$ on itself is transitive the value
of any element $\phi \in V_{0,j}$ is determined by its
value at the identity. Since the possible values at 
the identity are parametrised by  $\bC^{2j+1}$ we see explicitly that 
$V_{0,j}$ is $(2j+1)$-dimensional.  We further define  a  basis 
$\{\phi^{(m)}\}_{m=-j,-j+1,  ...,j-1,j}$ consisting of 
those elements of $V_{0,j}$ which reduce to
the canonical basis of $\bC^{2j+1}$  at the identity: 
$\phi^{(m)}_s(E) = \delta_{ms}$.
It then follows that 
\eqn\basis{
\phi^{(m)}_s(Q) = D^j_{sm}(Q^{-1}) = D^{j*}_{ms}(Q). 
}
By now it will be apparent to the reader that 
 Wigner functions are omnipresent in the representation theory
of the Euclidean group, where they play a number of different roles.
Here we find them as the components of $\bC^{2j+1}$-valued 
basis functions
of $V_{0,j}$. Note in particular that under the $SU(2)$ transformation
\acton\ the components of the vector-valued functions $\phi^{(m)}$
do not get mixed:
\eqn\acc{
(A\circ\phi^{(m)})_s(Q)=\phi^{(m)}_s(A^{-1}Q) = \sum_{l=-j}^j D^j_{lm}(A)
\phi^{(l)}_s(Q).
} 
Thus, for any fixed $s$  in the allowed range 
the component functions $\phi^{(m)}_s(Q)= D^{j*}_{ms}(Q)$,
$m\in\{-j,-j+1, ..., j-1,j\}$ span an equally valid carrier
space of the spin $j$ representation of $SU(2)$.

In order to complete our account of  the representation theory of 
$U(2)\semidir \bR^4$,  we need to  combine the above representations
with representations of $U(1) \semidir \bR$. Thinking about the latter
from the point of view of induced representation may seem unnecessarily
complicated but it is useful for a unified view. The $U(1)$ action 
on $\bR$ (by conjugation) leaves every point fixed, so all  orbits
 are  trivial and consist of 
a real number $K\in \hat \bR$. The centraliser is always the whole 
of $U(1)$, so centraliser representations are labelled by a single 
integer $N$. All UIR's are one-dimensional and given by
\eqn\abn{
 v_{K,N}= \bigl\{\phi:U(1) \rightarrow \bC\big \vert 
\phi(\chi +\xi)= e^{iN\xi}\phi(\chi)\,\, \forall \chi,\xi \in [0,2\pi)
\bigl\}.
}
Clearly this is just the one-dimensional space  spanned by
 the function $\phi_N(\chi)=e^{iN\chi}$. The action of an element $(\xi,a)\in
U(1)\semidir\bR$ on this function is
\eqn\silly{
 (\xi,a)\circ\phi_N(\chi) = e^{iKa}\phi_N(\chi-\xi).
}
The representations of $SU(2) \semidir \bR^3$ and of $U(1)\semidir\bR$
can now be combined to representations of $U(2)\semidir \bR^4$
by taking tensor products of individual representations, but respecting
the $\bZ_2$  conditions on $N$ and $j$ or $N$ and $s$ outlined earlier.
Thus carrier spaces of  UIR's of $U(2)\semidir \bR^4$ are of the form
\eqn\comnbi{
V_{K,k;N,s} = v_{K,N}\otimes V_{k,s}
}
with $k>0$ and  $N+2s$ even, or
\eqn\conbi{
V_{K,0;N,j} =  v_{K,N} \otimes V_{0,j}
}
with $N+2j$ even. We will use the unifying notation $V_{K,k;N,n}$
for these representations, with $n$ standing for the $U(1)$
representation label $s\in {1\over 2}\bZ$ if $k> 0$ 
  for the $SU(2)$
representation label $j\in \{0,{1\over 2},1, ...\}$ if $k=0$.

We now come to the promised identification of dyonic quantum 
 states of $SU(3)$ monopoles 
with elements of  $U(2)\semidir \bR^4$  representations,
 partly already anticipated by our notation.
Our proposal is to identify the magnetic 
charges  with the eigenvalues of the translation generators $P$
and $p_1,p_2,p_3$
of $U(2)\semidir \bR^4$, and the electric charges with the 
representations of the centraliser subgroups of $U(2)$.
More precisely we identify the topological magnetic charge
with the representation label  $K$ and the
magnitude of the  non-abelian magnetic
charge with the  representation label $k$. Then we associate
the dyonic states discussed earlier with the representation
spaces of  $U(2)\semidir \bR^4$ as follows.
For $k>0$, the  realisation of   states $|K,\bk; N, s\rangle$
 as wavefunctions on the moduli 
space \helss\ and \helsss\  shows  that they  can be 
written as infinite sums of elements of  $V_{K,k;N,s}$.
 In fact these sums do not converge, but since these 
states  are 
eigenstates of the translation generators $p_1,p_2$ and $p_3$, 
their  non-normalisability 
is the familiar property of of momentum eigenstates  in 
 quantum mechanics on $\bR^3$. Keeping that proviso in mind
we write
\eqn\reult{
|K,\bk; N, s\rangle \in V_{K,k;N,s}.
}
The dyonic states with $k=0$
can be interpreted as   
elements of the  representation space $V_{K,0;N,j}$
if we identify the label $s$ in the realisation of $|K,0; N, j,m \rangle$
 \utwost\ with 
the $s$-th component of elements of $V_{K,0;N,j}$: 
\eqn\biza{
\langle \chi',\alpha',\beta',\gamma'|K,0; N, j,m \rangle_s =
e^{iN\chi'}\phi^{(m)}_s(Q').
}
With this identification we can then also write
\eqn\reelt{
|K,0; N, j,m \rangle \in V_{K,0;N,j}.
}

One immediate question which arises after this identification
concerns the  quantisation of the magnetic charges. The eigenvalues
of the translation  generators are not naturally quantised, so 
by interpreting quantum states of magnetic monopoles 
as eigenstates of the translation operators we select a subset
of translation eigenstates by hand, namely those  with integer eigenvalues $K$
and half-integer values for the magnitude of $\bk$. We return to
the group theoretical interpretation of this quantisation 
in the final section of this paper.
 Now we impose the quantisation and press on, 
turning to the computation of the combination rules of dyonic states.
These can now be found  using 
the Clebsch-Gordan  coefficients of 
$U(2)\semidir \bR^4$.
The interesting part of this is again the 
(double cover)
of the Euclidean group, whose Clebsch-Gordan coefficients 
can be found in the literature see e.g. \cgcoeff. Their calculation
is lengthy (remarkably we could not find them  in 
any group theory textbook) because of the subtleties of combining
representations with different orbits and centraliser representations.
To give a flavour of the subject we note the following Clebsch-Gordan
series for the tensor product of two representations. Multiplying 
 two representation  with non-vanishing
orbit sizes $k_1$ and $k_2$ one finds:
\eqn\cgser{
V_{k_1,s_1} \otimes V_{k_2,s_2} = \int_{|k_1-k_2|}^{k_1+k_2} dk
\sum_{t=-\infty}^\infty \,V_{k,s_1+s_2+t},
}
where $k=0$ is allowed on the right hand side; in that case  the
sum over
$s_1+s_2+t$ is a sum over $SU(2)$ representations and should be restricted
to positive integers. 
Physically one may think of  this formula in terms of combining plane
waves with wave vectors $\bk_1$ and $\bk_2$ and helicities
$s_1$ and $s_2$. If the magnitudes of the wave vectors are fixed
to be $k_1$ and $k_2$ the combined plane wave may have wave vectors
with length $k$ varying between $|k_1-k_2|$ and $k_1+k_2$ and 
a helicity which is integer or half-odd integer depending on
the values of $s_1$ and $s_2$ but otherwise arbitrary.
If $k_1=k>0$ and $k_2=0$, the tensor product splits 
into a finite sum of irreducible representations:
\eqn\cgserr{
V_{k,s} \otimes V_{0,j} = \bigoplus_{n=-j}^j V_{k,s + n}.
}
Finally the case $k_1=k_2=0$ reproduces the familiar $SU(2)$
Clebsch-Gordan series:
\eqn\cgserr{
V_{0,j_1} \otimes V_{0,j_2} = \bigoplus_{j=|j_1-j_2|}^{j_1+j_2}
 V_{0,j}.
}

We are actually interested in more detailed information, 
namely the Clebsch-Gordan coefficients which specify
the relation between states.
On the other hand we only need to consider very special
states, namely those singled out by the Dirac quantisation
condition on the magnetic charge. This condition should
be imposed on both  the states to be multiplied and on the 
resulting product state, and this drastically  reduces 
the number of tensor products we need to consider.
Here we  have only treated quantum states of 
monopoles of  topological magnetic charge 1 and 2, but 
we propose as a general requirement that the  non-abelian
magnetic charges $\bk_1$ and $\bk_2$ of two states to be 
multiplied must be parallel or anti-parallel. 
It is clear that then all states obtained as the tensor
of two states which individually satisfy the Dirac condition
will also satisfy that condition.

Starting with quantum states of topological charge one
monopoles $ |1,\bk_1; N_1, s_1\rangle$ and $|1,\bk_2;N_2, s_2\rangle$
for example, where $k_1=k_2 =1/2$ and $N_1+2s_1 =N_2+2s_2 =0$,
 we  may thus  only combine
them if  either
$\bk_1 = \bk_2$ or   $\bk_1 = -\bk_2$.  
In the former case the tensor product
is
\eqn\tenab{
 |1,\bk;N_1, s_1\rangle \otimes  |1,\bk;N_2, s_2\rangle
=  \delta^2(0) |2, 2\bk;N_1+N_2, s_1+s_2\rangle,
} 
where the infinite factor $\delta^2(0)$  arises because we are
working with non-normalisable states. The equations \tenab\
is essentially the combination rule for monopoles
with abelian magnetic charges: the direction of  $\bk$
does not come into play in any interesting way. 
This agrees with the fact that   
 the quantum states
in \tenab\ can all be realised as wavefunctions on the 
smallest stratum of the relevant moduli space and that these
strata are fibred of the magnetic orbit, with fibres being isomorphic to
 $SU(2)$ monopole moduli spaces. The combination rule \tenab\
is the combination rule of $SU(2)$ monopoles in a fixed fibre. 
The possibility  of anti-parallel  non-abelian magnetic 
charges is much more interesting. Now the tensor product is
\eqn\tenabc{\eqalign{
 |1,\bk;N_1, s_1\rangle& \otimes |1, -\bk;N_2, s_2\rangle \cr
& = \delta^2(0) \,\sum_{j=|s_1-s_2|}^{\infty}\sum_{m=-j}^j \sqrt{2j+1}\,
d^j_{m(s_1-s_2)}(\beta)e^{-im\alpha} \,\,
|2,0;N_1+N_2,j,m\rangle_{s_1-s_2},
}}
where $(\alpha,\beta)$ are again the angles determining the direction
of $\bk$ as in \magvec. Here  the  representation theory  of 
$U(2) \semidir \bR^4$  has entered in a non-trivial way
and has solved one of the main puzzles of non-abelian
dyon physics,  namely 
how to combine two dyons  carrying non-abelian
magnetic charge and $U(1)\times U(1)$ electric charge into a dyon carrying
only topological magnetic charge and a $U(2)$ representation.
Particularly we now also see how  to decompose 
a quantum state $|2,0;N,j,m\rangle$ of a charge two monopole
 in terms of tensor product states of charge one monopoles
\eqn\pula{\eqalign{
\delta^2(0)|2,0;N,j,m\rangle_s =& \cr 
\int \sin \beta d\beta d\alpha \,\, & {\sqrt{2j+1} \over  4\pi}
d^{j*}_{ms}(\beta) e^{im\alpha}\,
|1,\bk;N_1, s_1\rangle \otimes  |1, -\bk;N_2, s_2\rangle,
}}
with the condition $N_1+N_2 =N$ and $s_1-s_2=s$.
Note that {\it all} magnetic directions are needed on the right
hand side. If we had only considered  single monopole states
with a particular magnetic directions we would not be able
to make sense of general charge two monopole quantum states
in terms of tensor product states of charge one monopoles.

We have not discussed monopoles of topological charge  three
here, but for completeness we also write down the rule for
combining a quantum state of a charge one monopole with 
a quantum state of a charge two monopole in the large stratum.
The former carries $U(1)$ electric charge and latter $U(2)$
electric charge, but within the representation theory of 
$U(2) \semidir \bR^4$  there is no problem in combining such
states. The answer is
\eqn\tenabcd{
 |1,\bk;N_1, s_1\rangle \otimes |2,0;N_2,j,m\rangle_{s_2}
={\sqrt{2j+1} \over 4 \pi}
  e^{im\alpha} d^{j*}_{ms_2} |3,\bk;N_1+N_2, s_1+s_2\rangle,
}
with $\bk$ of length $1/2$ and direction given by $(\alpha,\beta)$.

Having demonstrated the use of 
interpreting  dyonic quantum states of $SU(3)$ monopoles 
as elements of $U(2) \semidir \bR^4$
representations
we end this section by highlighting two restrictions
which are dictated by the physics of monopoles but which
are not naturally part of $U(2) \semidir \bR^4$
representation theory.
The first  restriction is the superselection rule that 
states with different (topological or non-abelian) magnetic
charge may not be superimposed.
The second is a consequence of the Dirac condition and  
is the restriction on which
states may be multiplied  in tensor products.

\newsec{BPS quantum  states and S-duality}

Maximally broken 
${\cal N}=4$ supersymmetric Yang-Mills theory is widely believed to 
enjoy exact  invariance under S-duality transformations. In general
but precise terms this statement means the following.
For  a given gauge group $G$ and symmetry breaking
to a group $H$ consider the two real parameters which uniquely 
characterise the action of ${\cal N}=4$ supersymmetric Yang-Mills theory,
namely the coupling constant $e$ and the $\theta$-angle. 
Then construct the 
  complex number  $\tau = \theta/2 \pi + 4 \pi i /e ^2$
in the upper half plane. An  S-duality transformation is a modular
transformation on $\tau$
\eqn\sdual{
\tau \rightarrow {q\tau -r  \over -p \tau + s},
}
where $M = \left( \matrix{ q & -r \cr -p & s} \right) \in SL(2,\bZ)$,
together with a suitable $ SL(2,\bZ)$ action  on the BPS
 states of the quantum theory (to be defined presently).
If  $H$ is a  maximal torus of $G$, the 
 BPS states  are  dyonic states  characterised by 
$R=$rank$(H)$=rank$(G)$ pairs of 
integers $(m_l,n_l)$, $l=1, ... ,R$, giving  the magnetic  and 
electric  charges respectively \abduality.  Under S-duality these states
transform as 
\eqn\sdualst{
(m_l, n_l) \rightarrow 
(m_l, n_l) M^{-1}
}
for all  $l=1, ..., R$. In particular the electric-magnetic
duality operation originally considered by Montonen and Olive \MO\
is  given by the matrix
\eqn\emdu{\eqalign{
M = \left( \matrix{ 0 & 1 \cr -1 & 0} \right),
}}
which exchanges strong with weak coupling and  electric with
magnetic charges.

Little is known about duality in theories with 
non-abelian unbroken gauge symmetry.
There exists  a conjecture, due to Goddard, Nuyts and Olive \GNO\
according to which 
Yang-Mills-Higgs  theory with gauge group $G$ broken to $H$ has 
a dual description at strong coupling in terms of weakly coupled
Yang-Mills-Higgs theory with  a dual gauge group $\tilde G$
broken to $\tilde H$.
The GNO conjecture interprets the non-abelian
monopole charges, after rotation into the Cartan subalgebra, 
as labels of irreducible interpretations of 
the dual group $\tilde H$. The true invariance group 
 of Yang-Mills-Higgs  theory with unbroken gauge group $H$
would, according to the GNO conjecture, be the product 
$H\times  \tilde H$ of an ``electric'' and a ``magnetic''
 version of the unbroken gauge group. 
In particular the GNO conjecture would imply that dyonic
states   fall into  representations of that product group.
However, here we have seen that dyonic quantum states of 
monopoles in minimally broken $SU(3)$ Yang-Mills-Higgs theory
have correlated magnetic and electric properties,
with the magnetic orbit determining the  part of the unbroken
gauge group with respect to which the dyons carry electric charge.
This correlation is not accounted for  by the representation
theory of the GNO group $U(2)\times \tilde U(2)$, but, as we have seen,
it is captured perfectly by the representation theory of the semi-direct
product $U(2)\semidir \bR^4 $.  In particular the Clebsch-Gordan
coefficients of  $U(2)\semidir \bR^4 $ lead to combination rules
for dyonic states which are consistent with their realisation 
as wavefunctions on the moduli spaces.

The goal of this section is to look  at the implications of 
 our insights into the dyonic spectrum
for S-duality in Yang-Mills theory with    non-abelian
unbroken gauge symmetry.  
We will consider   transformation rules for dyonic
states  which generalise the abelian
rules \sdualst\   to  the non-abelian case with 
minimal modifications. In particular we thus consider transformation
rules which map  the dyonic states of minimally broken
$SU(3)$  Yang-Mills-Higgs theory  into dyonic states of  the same theory 
with dual coupling \sdual. We are aware that this may well be
too restrictive. In particular we feel that 
one needs a better understanding of  duality in the massless sector
of the ${\cal N}=4$ supersymmetric version of the theory
before one can gain a full understanding of duality. 
However, the massless sector 
has essentially the same particle
contents as unbroken ${\cal N}=4$ supersymmetric $U(2)$ Yang-Mills theory, 
whose dual formulation is not known \lozano. We do not attempt
to solve this problem here and focus on  the massive particles instead. 
As we shall see, 
there is a natural 
$SL(2,\bZ)$-duality action on the massive dyonic states,
whose study is instructive.   We are particularly interested 
in BPS states, which are (in general)
 dyonic states 
in some  representation $ V_{K,k;N,n}$ of $U(2)\semidir \bR^4$
whose  energy equals the BPS bound
\eqn\bpsen{
E_{\footnotefont{BPS}}= ve|N + \tau K|.
}
where $v$ is the magnitude \vev\ of the 
vacuum expectation value of the Higgs field.

The starting point for the transformation rule we want to consider
 is the  observation of the previous section that dyonic quantum
states fit into certain representations of $U(2) \semidir \bR^4$.
This generalises in the obvious way to  
dyonic states in Yang-Mills-Higgs theory with general unbroken
gauge group $H$: these states   fall into  certain representations of
 $H \semidir \bR^D$ ($D=$dim$(H)$. 
Thus (as we shall explain in more detail in \SaBe)
 they are labelled by $R$ quantised orbit parameters
and by $R$ labels specifying the representation of the 
centraliser group associated to the orbit (which  always contains
a  maximal torus and  therefore has the same rank as $H$).
This is also (trivially) true in the maximal symmetry breaking case,
where the orbits are always just points in  a $R$-dimensional
lattice and the centraliser is always the whole of $H$.
In that case, as we saw above, electric-magnetic duality acts
by exchanging orbit labels with centraliser representation labels.
In  this formulation  the  electric-magnetic duality
transformation rule naturally generalises 
to the situation  where  $H$ is non-abelian.
Again we  focus on 
the case $G=SU(3)$ and $H=U(2)$. 
  
In that case we have seen that magnetic orbits are labelled by
the pair $(K,k)$, with $K\in \bZ$ and $0\leq  k \leq K/2$ 
integer or half-odd integer depending on whether $K$ is even or odd,
which may be summarised as the requirement that  $K+2k$  is even.
Centraliser representations  are labelled 
by  the pair $(N,n)$, where $N\in \bZ$ 
specifies a $U(1)$ representation and  $n$ defines a centraliser
representation as specified  after
\conbi.  Again we have a $\bZ_2$ 
condition relating $N$ and $n$, namely that $N+2n$  is even.
The S-duality action we want to 
consider involves the $SL(2,\bZ)$ action on the $\tau$-parameter \sdual\
together with the following action  on the representation labels:
\eqn\result{\eqalign{
(K,N) & \rightarrow
(K,N) M^{-1}  \cr
(k,n) &\rightarrow 
(k,n) M^{-1}.
}}
Recall that  in the abelian
case, where all representations are one-dimensional, 
there is either no or a unique BPS state (or, more precisely, a unique
short ${\cal N}=4$ supermultiplet) in the 
 representation labelled by  $(m_l,n_l)$. 
Thus there is no difference between  an $SL(2,\bZ)$
action on  representations and an $SL(2,\bZ)$
action on states. In the non-abelian case, however, we have
to decide whether we consider  the action of  S-duality  on 
the entire carrier space of  representations  or
 just on  certain states in those spaces.
A moment's thought shows that one cannot expect  the carrier spaces of the 
representations related by  \result\ to be ``physically equivalent'':
for example the purely electric representation $V_{0,0;1,1/2}$
is two-dimensional whereas the purely magnetic representation 
$V_{1,1/2;0,0}$ is infinite-dimensional. On may hope that 
the BPS condition will select the same number of states 
in  all representations related by \result, but this
is not case as we shall see. We  postpone a full discussion
of this puzzle until the end of this section. 
Here  we  note that the alternative route, that of 
considering the action of S-duality on  individual  
states in the spaces $V_{K,k;N,n}$, can be implemented quite easily.
 The idea is to   select a unique
state in each  carrier space $V_{K,k;N,n}$ by a ``natural''
condition (which  necessarily breaks the 
$U(2)\semidir \bR^4$ invariance of the carrier spaces maximally).
One can do this, for example,   by picking a  unit vector
$\hat \bk$ and in each  $V_{K,k;N,n}$ select that state which is
the eigenstate of the $SU(2)$  generator  $\hat \bk \cd \bI $ with
maximal (or minimal)  eigenvalue. If $k=0$ this condition selects
the state in the spin  $j$ representation of $SU(2)$ with spin
$j$ along the $\hat \bk$ axis; if $k>0$ it selects the state
$|K,k \hat \bk;N,n\rangle$ (provided  $n\geq 0$). 
While this condition may appear {\it ad hoc} we will see at the 
end of this section 
that it has played an important role in earlier discussions
of electric-magnetic duality.

Consider now the $SL(2,\bZ)$ orbit of the 
  massive $W$-bosons of  the theory on which previous 
discussions of electric-magnetic duality have traditionally
focused.
The massive 
$W$-bosons form a doublet under $SU(2)$ and carry one unit of $U(1)$
charge, so they belong to the representation $(K,k) = (0,0)$
and $(N,n)=(1,1/2)$. Under the $SL(2,\bZ)$ transformation \result\
 this is mapped to
\eqn\sorbet{
(K,k)= p(1,{\textstyle{1 \over 2}}) \qquad (N,n) = 
q(1,{\textstyle{1 \over 2}}),
}
for relatively prime integers $p$ and $q$. These states 
carry magnetic charge, and at weak coupling should
be visible semi-classically. More precisely,
since $k=K/2$,  they
should emerge as dyonic states in the smallest strata of the
moduli spaces we have described. Since these moduli spaces
consist of  embedded $SU(2)$ monopoles we can
hope to deduce the prediction of S-duality in this case from 
the S-duality properties of 
the  $SU(2)$ theory. This is in fact what we will 
do, but since our computation needs to be carried out
in the ${\cal N}=4$ supersymmetric version of the purely bosonic theory
we have described so far, we first need to explain how ${\cal N}=4$
supersymmetry is implemented in our quantisation scheme.

The general procedure for implementing supersymmetry in the 
collective coordinate or moduli space quantisation of monopole
dynamics is explained in \susymo\ in the context of 
$SU(2)$ monopoles. For ${\cal N}=4$ supersymmetry this procedure
leads to the Hilbert space of states being the space of 
all square-integrable  real differential forms on the 
moduli space and the Hamiltonian being the covariant Laplacian.
An important consistency requirement is that the metric on the 
moduli space is hyperk\"ahler. 
The arguments given in \susymo\ apply without essential changes
to the largest strata of the moduli spaces we have described,
i.e. those labelled by  $K$ even and $k=0$. These spaces
have dimension  $6K$ (which is thus divisible by four) and are 
equipped with smooth hyperk\"ahler metrics. All the other strata,
however,  have dimensions which are even but not divisible by four,
indicating clearly that the standard procedure needs to be 
amended.  In fact the  required changes follow naturally from
the magnetic superselection rules introduced in the previous
section. Namely, the Hilbert spaces ${\cal H}_{K,\bk}$ introduced after
\clebv\ can be extended to accommodate ${\cal N}=4$ supersymmetry in the 
standard fashion. Their elements are functions on the fibres
$M_{K,\bk}$ of the fibration \fibre\ and these spaces 
do have dimensions which are multiples of four. In the case
$k=K/2$  they are actually isomorphic to $SU(2)$ monopole
moduli spaces and thus in particular equipped with hyperk\"ahler
metrics. It has not been shown rigorously that  for $k< K/2$
these spaces also have smooth hyperk\"ahler metrics, but
non-rigorous physical arguments  suggest that they do.
Here we are in any case mainly interested in the case $k=K/2$,
so we fix this condition from now on.

Then we  define the space ${\cal H}_{K,\bk}^*$
of square-integrable real differential forms on $M_{K,\bk}$.
The quantum Hamiltonian  is the Laplacian on $M_{K,\bk}$, 
which equals, up to scaling, the Laplacian on the $SU(2)$
monopole moduli  space $M_K^{\footnotefont{SU(2)}}$.
The eigenstates of that Hamiltonian which saturate the 
BPS bound, however, were conjectured by Sen \sen\
to exist, and be unique,  for all $(K,N) =(p,q)$ relatively prime integers.
 There is now much support for the 
validity of the Sen conjecture \sen, \segal, and 
assuming it we deduce the existence of a unique dyonic BPS
state   for each value  of the non-abelian  magnetic charge $\bk$
in the $U(2) \semidir \bR^4$ representation $V_{K,K/2;N,N/2}$
with $(K,N) =(p,q)$ relatively prime integers. Thus we indeed  find  
the BPS states on the $SL(2,\bZ)$ orbit of the massive $W$-boson states, but
we encounter the puzzle anticipated earlier in this section:
whereas the massive $W$-bosons fill the two-dimensional representation
 $V_{0,0;1,1/2}$, the dyonic BPS states fill the 
infinite-dimensional representations $V_{K,K/2;N,N/2}$.  

In earlier discussion of duality e.g. in \GNO\
 this ``dimensionality 
puzzle'' is not seen because only vector magnetic charges  on the orbit of 
a given vector magnetic charge under the action of the Weyl group
of  the unbroken gauge group are considered. It follows from
general group theory that the number of points on the Weyl
orbit of the vector magnetic charge of a topological charge one
monopole is equal to the dimension of the fundamental 
representation of the unbroken gauge group. Thus, with 
this counting procedure  the number
of massive $W$-boson states agrees with the number of topological
charge one monopoles. This observation is in fact a key
ingredient in the formulation of the  GNO conjecture
in \GNO. Here, however, we have seen that the full
magnetic orbit is essential for understanding dyonic states
and their interaction, and that any restriction to a subset
of the orbit is artificial and will miss some of the physics 
of dyons. From our point of view the restriction to 
Weyl orbits is equivalent to choosing a direction $\hat \bk$
and restricting attention to states which are eigenstates of 
the operator $\hat\bk \cd \bI$ introduced earlier in this section.
Thus GNO duality in the theory considered here would, in
our language, amount to choosing $\hat\bk = (0,0,1)$ and selecting
the eigenstates of $I_3$ with maximal and minimal eigenvalues.
This condition 
selects the usual basis of massive $W$-boson states  in the purely electric 
sector $V_{0,0;1,1/2}$  while in the magnetic sector $V_{1,1/2;0,0}$ 
it picks out two  monopoles with magnetic charges on
the same  Weyl orbit 
 (the Weyl group of $U(2)$ is
 the permutation group $S_2$ of two elements  and 
acts on the magnetic orbits we have defined by
sending a point to its antipodal point).
Such a choice can be made,  but it is {\it ad hoc} and breaks
the symmetry of the theory in an artificial way.

We end this section with two comments.
The first concerns the place  of Weyl orbits in our description of dyons.
Although we do not select any {\it particular} Weyl orbit in our scheme,
 Weyl orbits do play an important role. 
Dyonic states with non-abelian magnetic charges in the same Weyl orbit
have  parallel or anti-parallel non-abelian
magnetic charges  and are thus  precisely those 
states in $V_{K,k;N,s}$ which  may be multiplied in  a tensor product.
 Thus, although the representation
space $V_{K,k;N,s}$ is infinite-dimensional for $k>0$, a given 
state  can interact consistently with at most two 
other states in the same representation. 
This is intriguingly reminiscent of the role Weyl orbits play
in the physics of classical non-abelian electric charges.
In \Corrigan\ points on  Weyl orbits were  shown to correspond
one-to-one to the relative orientations which two non-abelian electric
point charges may have if the non-abelian
electric field they produce is to be static.
These observations are not  a sufficient reason to assign 
any dimension other
than infinity to the carrier spaces $V_{K,k;N,s}$ ($k>0$) but 
they do show that counting degrees of freedom for particles 
carrying non-abelian charges is a subtle business.

The second comment concerns  the paper \Swansea. 
In that paper,  a regularisation procedure was used 
to make motion on (in our 
language) the  magnetic orbit  dynamically possible and  a supersymmetric
quantisation scheme was adopted which allowed forms on the magnetic
orbit as physical  states. In that scheme 
 computing the degeneracies
of magnetic states is equivalent to counting harmonic forms on
the magnetic orbit, i.e. to  computing its  Euler characteristic.
However,  the Euler characteristic of the coset
spaces considered in 
\Swansea\ is equal to the number of points in the orbit
under the Weyl group of the unbroken gauge
group of any point in the coset. We relegate a discussion and proof
(which is a simple application of  Morse theory)
of this statement for general gauge groups to \SaBe,
but in the theory  considered in this paper  the  validity is 
easily demonstrated. Here the relevant coset is the magnetic orbit
whose  Euler characteristic is two if $k>0$
and one  if $k=0$. Similarly the  orbits of the Weyl group acting on
the magnetic orbit have order two if $k>0$ and order one
if  $k=0$.
Thus, the  counting procedure
for magnetic states described  in \Swansea\
also boils down to  determining
the order of Weyl orbits.

\newsec{Discussion and outlook}

The mathematical structure of the  monopole moduli spaces
described in this paper singles out certain coordinates,
called magnetic orbits here, and only allows 
smooth isometric actions of a  subgroup of the unbroken
gauge group,  namely the  centraliser group associated to 
the orbit. As a result magnetic orbits and  electric centraliser
representations turn out to be the natural labels of dyonic
quantum states in minimally broken $SU(3)$
 Yang-Mills-Higgs theory. Noting that 
orbits and centraliser representations  are also the 
attributes of representations of the semi-direct product
$U(2) \semidir \bR^4$ we have proposed to interpret
dyonic states as elements of such representations. 
It is worth stressing again that this group is not a symmetry
of the classical theory and that in particular the 
magnetic part $\bR^4$  does not act 
on either the classical configuration space  or the moduli spaces.
Nonetheless we have seen that the wavefunctions on the moduli space
are acted upon by the  full group $U(2) \semidir \bR^4$.
Thus, although one usually thinks of the electric excitations
as quantum effects and the magnetic properties as classical,
we find that in our scheme the magnetic properties are also
encoded in quantum wavefunctions. Furthermore, the interpretation
of dyonic states as elements of representations of $U(2) \semidir \bR^4$  
leads to combination rules of dyonic states 
which are consistent with the realisation of dyonic
states as wavefunctions on  the appropriate moduli spaces.

However, while  thinking of dyonic states as  elements of 
$U(2) \semidir \bR^4$ representations turns out to be
very fruitful, it also raises a number of questions, the most
important of 
of which  concerns  the 
quantisation of  the magnetic orbit sizes: from 
the point of view of $U(2) \semidir \bR^4$ representations
this is an artificial condition, while in monopole physics
it is one of the most basic facts. Ideally we would like
the algebraic object which classifies dyonic quantum
states to incorporate quantised magnetic orbit sizes 
(or equivalently the  Dirac quantisation condition)
as a basic ingredient. It remains a challenge to find that
algebraic object, which we expect to be closely related to
semi-direct product groups.

A second question concerns the  S-duality properties of 
the dyonic spectrum. We have seen that it is possible to 
define  an $SL(2,\bZ)$-action on the dyonic states of the theory, with
the  electric-magnetic duality operation exchanging 
labels of  the  magnetic orbits
 with  labels of the  centraliser representations. 
With this definition  
the basic massive $W$-boson states are seen to be part 
of an $SL(2,\bZ)$-orbit whose other elements are    
dyonic BPS states  which  can be found semi-classically
as quantum states on the smallest strata of monopole moduli spaces. 
However, in interpreting these BPS states as elements of
 $U(2) \semidir \bR^4$ representations we are forced to conclude 
that S-duality relates states  belonging to  
$U(2) \semidir \bR^4$ representations of different 
dimensions. We expect that 
a solution of this ``dimensionality puzzle'' will require a deeper
understanding of S-duality in supersymmetric Yang-Mills theory
with non-abelian unbroken gauge symmetry.

To end, we want to stress a link to physics
 in (2+1) dimensions. In \BDP\ it was shown that 
dyonic quantum states of vortices  in  (2+1)-dimensional
gauge theories  with discrete unbroken gauge groups 
are labelled by  magnetic fluxes together with a 
centraliser representation of the unbroken gauge group.
There the relevant algebraic object whose 
representations classify the dyonic states is the so-called
quantum double $D(H)$ of the unbroken gauge group $H$, which
by definition is the  tensor product of the algebra of functions
on the group with the group algebra. Dyons in two spatial 
dimensions have distance-independent topological interactions
(Aharanov-Bohm scattering, flux metamorphosis) and 
remarkably these can be deduced from the algebraic structure 
of $D(H)$. More recently quantum doubles have also been 
constructed for continuous groups, such as $SU(2)$ \MK\
and were found to have a number of structural similarities
to semi-direct product groups of the type discussed here,
with the function algebra in the quantum double being
related to the translation part of the semi-direct product
groups.  The intriguing relation between these two algebraic
objects deserves to be studied further. The important
lesson one learns from studying quantum states of both
vortices with non-abelian flux and monopoles with non-abelian
magnetic charge is that consistent algebraic classifications
of these states require that one treats magnetic and electric 
properties as interdependent aspects of one algebraic
object.

\ack{

We are grateful to  Robbert Dijkgraaf, Erwyn van der Meer, Nathalie Muller, 
Herman Verlinde and Mark de Wild Propitius  for discussions.
BJS also thanks 
 Conor Houghton, Nick Dorey, Patrick Irwin, 
Stuart Jarvis, Klaas  Landsman, 
Nick Manton, Jae-Suk Park  and especially Andrew Dancer for discussions,
is grateful to the Isaac Newton Institute for hospitality  during
a two-week visit in January 1997   
and  acknowledges financial support through a  Pionier Fund of 
the Nederlandse Organisatie voor Wetenschappelijk Onderzoek (NWO). 
}

\listrefs
\bye